\documentclass[11pt]{book}

\usepackage{amssymb}
\usepackage{latexsym}
\usepackage{theorem}

\newcommand{\be}{\begin{equation}}
\newcommand{\ee}{\end{equation}} 
\newcommand{\bea}{\begin{eqnarray}}
\newcommand{\beann}{\begin{eqnarray*}}
\newcommand{\eea}{\end{eqnarray}}
\newcommand{\eeann}{\end{eqnarray*}}
\newcommand{\ov}{\overline}
\newcommand{\la}{\langle}
\newcommand{\ra}{\rangle}
\newcommand{\dz}{\frac{d}{dz}}
\newcommand{\fn}{\textstyle {\circ \atop \circ}}
\newcommand{\D}{\displaystyle}
 
\theoremstyle{break}

\newtheorem{p1}{Proposition}[chapter]
\newtheorem{l1}{Lemma}[chapter]

\newtheorem{thp1}[p1]{Theorem}

\newtheorem{pp1}[p1]{Proposition}

\newtheorem{lp1}[p1]{Lemma}
\newtheorem{ll1}[l1]{Lemma}

\begin{document}

\thispagestyle{empty}
\vspace*{0cm}
\begin{center}
{\huge \bf Vertex algebras, Lie algebras and \\[0.5cm]
           superstrings } \\[6cm]
{\large \bf Dissertation \\
zur Erlangung des Doktorgrades \\ 
des Fachbereichs Physik \\
der Universit\"at Hamburg} \\[2cm]
{\large vorgelegt von \\
Nils R. Scheithauer \\
aus Marburg\\[2cm]
Hamburg \\
1997}
\end{center}

\pagenumbering{roman}
\vspace*{\fill}
\begin{tabular}{lcl}
Gutachter der Dissertation:  & $\quad $ & Prof. Dr. H. Nicolai \\
                             &          & Prof. Dr. G. Mack \\
                             &          &  \\
Gutachter der Disputation:   &          & Prof. Dr. H. Nicolai \\
                             &          & Prof. Dr. K. Fredenhagen \\
                             &          &  \\
Datum der Disputation:       &          &  23.1.98\\
                             &          &  \\
Dekan des Fachbereichs       &          &  \\
Physik und Vorsitzender      &          &  \\
des Promotionsausschusses:   &          & Prof. Dr. B. Kramer
\end{tabular}

\newpage

\centerline{\Large\bf Abstract}

This work deals with certain vertex algebras and Lie algebras 
arising in superstring theory. 

We show that the Fock space of a compactified Neveu-Schwarz superstring, i.e. a Neveu-Schwarz superstring moving on a torus, carries the structure of a vertex superalgebra with a Neveu-Schwarz element. This implies that the physical states of such a string form a Lie algebra. The same is true for the GSO-projected states. The structure of these Lie algebras is investigated in detail. In particular there is a natural invariant form on them. In case that the torus has Lorentzian signature the quotient of these Lie algebras by the kernel of this form is a generalized Kac-Moody algebra. The roots can be easily described. If the dimension of space-time is smaller than or equal to $10$ we can even determine their multiplicities.\\[2cm]

\centerline{\Large\bf Zusammenfassung}

In dieser Arbeit werden gewisse Vertex-Algebren und Lie-Algebren, die in der Superstring-Theorie auftauchen, untersucht.

Es wird gezeigt, da{\ss} der Fockraum eines kompaktifizierten Neveu-Schwarz-Superstrings, d.h. eines Neveu-Schwarz-Superstrings, der sich auf einem Torus bewegt, die Struktur einer Vertex-Superalgebra mit Neveu-Schwarz-Element besitzt. Damit k\"onnen wir beweisen, da{\ss} die physikalischen Zust\"an\-de eines solchen Strings eine Lie-Algebra bilden. Dasselbe gilt f\"ur die Zust\"an\-de invariant unter der GSO-Projektion. Diese beiden Lie-Algebren werden genau untersucht. Es wird eine kanonische invariante Bilinearform auf diesen Lie-Algebren konstruiert. Hat der Torus lorentzsche Metrik, so ist der Quotient dieser Lie-Algebren mit dem Kern der Bilinearform eine verallgemeinerte Kac-Moody-Algebra. Die Wurzeln dieser Lie-Algebra lassen sich leicht angeben. Ist die Dimension des Torus kleiner oder gleich $10$, so lassen sich sogar deren Multiplizit\"aten berechnen.   

\tableofcontents

\chapter{Introduction}
\setcounter{page}{1}
\pagenumbering{arabic}

Lie algebras were introduced in mathematics 1888 by S. Lie to describe
continous groups infinitesimally. Since then they play an important role in many branches of mathematics and also in physics. 

The theory of finite-dimensional Lie algebras is nowadays well understood and one of the most beautiful theories in mathematics. The contrary is true for the theory of infinite-dimensional Lie algebras. Even if one only considers the class of Kac-Moody algebras there are many open problems. The main reason for this is that we do not know natural realizations of them apart from the finite-dimensional and the affine Kac-Moody algebras. On the other hand the big success of the latter is due to the isomorphism of affine algebras and current algebras. 
This can be interpreted in the way that the class of Kac-Moody algebras is too restricted as to contain further natural examples.

One solution to this problem was shown by R. E. Borcherds in a series of papers (\cite{Bo1}-\cite{Bo5}). He defined the bigger class of generalized Kac-Moody algebras, showed that most of the nice properties of Kac-Moody algebras still hold and gave some very interesting examples, a.o. the famous monster Lie algebra and the fake monster Lie algebra. In the construction of the examples string theory and vertex algebras came into play. 

Let us describe this in more detail.

Borcherds defined the notion of a vertex algebra with Virasoro element
and stated the following results:\\
1) The Fock space of a compactified bosonic string, i.e. 
a bosonic string moving on a torus, carries this structure.\\ 
2) The natural bilinear form on this vertex algebra is invariant.\\
He showed that the physical states $P_1$ of the bosonic string modulo the null states $L_{-1}P_0$ form a Lie algebra.\\
Then he defined the notion of a generalized Kac-Moody algebra and proved that 
$P_1/L_{-1}P_0$ devided by the kernel of the bilinear form is a generalized Kac-Moody algebra if the metric of the torus is Lorentzian. He gave the multiplicities for the cases $d\leq 26$.

In this work we develop similar ideas for the Neveu-Schwarz superstring.
Thereby we also prove some of Borcherds statements for which no proof has yet been published.

The work is organized as follows.
 
In the second chapter we list some facts about vertex superalgebras and prove a number of new results. In particular we show that the quotient space $G=G_{-\frac{1}{2}}\mbox{\sf P}_{\frac{1}{2}}/L_{-1}\mbox{\sf P}_0$ of a vertex superalgebra 
with Neveu-Schwarz element is a Lie algebra. Furthermore we develop a theory of invariant bilinear forms on vertex superalgebras and determine the space of such forms. As an easy consequence the bilinear form
defined by Borcherds in \cite{Bo1} does indeed have the claimed invariance property.

In the third chapter we define the vertex superalgebra of the compactified 
Neveu-Schwarz superstring and show that it contains a Neveu-Schwarz element. 
This gives us the Lie algebra $G$ of physical states of the compactified       Neveu-Schwarz superstring. We show that it contains a Kac-Moody algebra generated by the tachyon and the first excited states and give an upper bound for the multiplicities. 
Then we prove that the states invariant under the GSO-projection also form a
Lie algebra $G^+$. With the theorem on the existence of invariant forms 
we construct an invariant form  
on the Lie algebras $G$ and $G^+$. Using a characterization of generalized Kac-Moody algebras due to Borcherds we prove that in the case of Lorentzian 
space-time the quotients of these Lie algebras by the kernel of this bilinear form are generalized Kac-Moody algebras. Finally we give the roots of these algebras and determine their multiplicities in the cases $d\leq 10$.
The results in this chapter are all new.

Some ideas on the construction of generalized Kac-Moody superalgebras describing the full superstring moving on a torus and a fake monster superalgebra are presented in chapter 4.  

In the appendix we calculate the dimensions of physical subspaces since we need them to compute the multiplicities. Moreover we prove Borcherds result on the multiplicities. 

Chapter 2 and 3 of this work will appear in the "Journal of Algebra" (cf. \cite{NRS}).

\chapter{Vertex superalgebras}

In this chapter we recall some basic facts about vertex superalgebras. Furthermore we prove some new results that are crucial for the construction of the generalized Kac-Moody algebras. In particular we will show that the quotient space $G_{-\frac{1}{2}}\mbox{\sf P}_{\frac{1}{2}}/L_{-1}\mbox{\sf P}_0$ in a vertex superalgebra with a Neveu-Schwarz element is a Lie algebra and we determine the space of invariant forms on a vertex superalgebra.

For an introduction into vertex algebras and vertex superalgebras consider 
\cite{FLM},\cite{FHL} and \cite{K2}. Proofs which are not given here can be found there.

\section{Definition and some properties}

In the following we will work over the real numbers.

In a ${\mathbb Z}_2$-graded vectorspace $V=V_{\ov{0}}\oplus V_{\ov{1}}$ we put $|v|=0$ if $v\in V_{\ov{0}}$ and $|v|=1$ if $v\in V_{\ov{1}}$. Elements in $V_{\ov{0}}$ or $V_{\ov{1}}$ are called homogeneous. Whenever $|v|$ is written, it is to be understood that $v$ is homogeneous.

A vertex superalgebra is a ${\mathbb Z}_2$-graded vectorspace $V=V_{\ov{0}}\oplus V_{\ov{1}}$ equipped with an infinite number of products, written as $u_n v$ for $u,v  \in V,\, n\in {\mathbb Z}$, satisfying the following axioms
\begin{enumerate}
\item The products respect the grading, i.e. $|u_n v|=|u|+|v|$.
\item $u_n v=0$ for $n$ sufficiently large
\item There is an element $1\in V_{\ov{0}}$, called vacuum, such that $v_n 1=0$ for 
             $n\geq 0$ and $v_{-1}1=v$  
\item The Jacobi identity 
            \beann
            \lefteqn{\sum_{0\leq k} {m \choose k}(u_{l+k}v)_{m+n-k}w=} \\
            & &    \sum_{0\leq k} {l \choose k}(-1)^k \left\{
                   u_{m+l-k}(v_{n+k}w)-(-1)^l(-1)^{|u||v|}v_{n+l-k}(u_{m+k}w)
                    \right\}
            \eeann holds for all integers $l,m$ and $n$.
\end{enumerate}

$u_n$ can be considered as an endomorphism of $V$ mapping $v$ into $u_nv$. This gives another approach to vertex superalgebras by vertex operators $Y(u,z)=\sum u_n z^{-n-1}$ which are formal Laurent series in $z$ with coefficients in $End(V)$. We will use both notations.

One can show that the Jacobi identity is equivalent to the locality condition
\[ (z_1-z_2)^n[Y(u,z_1),Y(v,z_2)]_{\mp}=0 \qquad \mbox{for $n$ sufficiently large.} \]
This gives Kac' definition of a vertex superalgebra.

A vertex algebra (cf. \cite{Bo2}) is a vertex superalgebra with trivial odd part $V_{\ov{1}}$.

With $l=0$ the Jacobi identity yields the commutator formula
\be [u_m,v_n]_{\mp} = \sum_{0\leq k} {m \choose k}(u_{k}v)_{m+n-k} \label{comf}
\ee
where $[u_m,v_n]_{\mp} =u_mv_n-(-1)^{|u||v|}v_nu_m$
and with $m=0$ the associativity formula 
\be (u_{l}v)_{n}= \sum_{0\leq k} {l \choose k}(-1)^k \left\{
       u_{l-k}v_{n+k}-(-1)^l(-1)^{|u||v|}v_{n+l-k}u_{k} \right\}.
\ee
We define a linear operator $D$ on $V$ by $Dv=v_{-2}1$. Note that $D$ preserves the spaces $V_{\ov{0}}$ and $V_{\ov{1}}$.
The following proposition summarizes some properties of the products and the operator $D$.
\begin{p1}
Let $V$ be a vertex superalgebra. Then for $u,v\in V$ and $n\in {\mathbb Z}$ 
\bea
1_nv    &=& \delta_{n+1,0}v \label{eins} \\
(Dv)_n  &=& -nv_{n-1}       \label{df} \\
D(u_nv) &=& (Du)_nv+u_n(Dv) \label{DistDer} 
\eea
and for homogeneous $u$ and $v$
\be 
u_nv-(-1)^{n+1}(-1)^{|u||v|}v_nu = 
(-1)^{n+1}(-1)^{|u||v|}\sum_{1\leq k}(-1)^k\frac{D^k}{k!}(v_{n+k}u). 
\label{qs} \ee
\end{p1}

Eq. (\ref{DistDer}) means that $D$ is for each product an even derivation. It implies $D(1)=D(1_{-1}1)=1_{-1}(D1)+(D1)_{-1}1=2D(1)=0$. In terms of vertex operators the equations (\ref{df}) to (\ref{qs}) can be written  
\bea
Y(Dv,z)    &=& \dz Y(v,z) \\
{[}D,Y(v,z){]} &=& Y(Dv,z)    \\
Y(u,z)v    &=& (-1)^{|u||v|}e^{zD}Y(v,-z)u.
\eea
An ideal of a vertex superalgebra $V$ is a $D$-invariant subspace $J$ such that
$J=J_{\ov{0}}\oplus J_{\ov{1}}$ where $J_{\ov{\imath}}=J\cap V_{\ov{\imath}}$ and $v_nJ\subset J$ for all $v\in V$. $V$ is called simple if it contains only trivial ideals. A homomorphism of $V$ is a linear parity preserving map $\psi : V\rightarrow V$ with $\psi(u_nv)=(\psi u)_n(\psi v)$. We have
\begin{pp1}
Let $V$ be a simple vertex superalgebra and $v\neq 0$ an element in 
$V_{\ov{0}}$ such that $Dv=0$. Then $v=\lambda 1$ for some $\lambda 
\in {\mathbb R}$.
\end{pp1}
{\it Proof:} The map $v_{-1}: V\rightarrow V$ is a homomorphism of $V$ (cf. \cite{Li2}). Eq. (\ref{df}) implies $v_n=0$ for all $n\neq -1$. Then by the commutator formula (\ref{comf}) $v_{-1}$ commutes with all $u_n$. Since $v_{-1}v=(v_{-1}1)_{-1}(v_{-1}1)=
v_{-1}(1_{-1}1)=v$ the map $v_{-1}$ has an eigenvalue. The proposition now follows from Schur's lemma. \hspace*{\fill} $\Box$\\

Equation (\ref{qs}) shows that the products are not symmetric on $V$. But we have
\begin{pp1}
Let $V$ be a vertex superalgebra. Then $V/DV$ is a Lie superalgebra under 
$[\ov{u},\ov{v}]=\ov{u_0v},\, u,v\in V$ with even part $V_{\ov{0}}/DV_{\ov{0}}$ and odd part $V_{\ov{1}}/DV_{\ov{1}}$.  
\end{pp1}
{\it Proof:} From $DV_{\ov{\imath}}\subset V_{\ov{\imath}},\, i=0,1$ it follows $DV\cap V_{\ov{\imath}}=DV_{\ov{\imath}},\, i=0,1$ and 
\[ V/DV=V_{\ov{0}}/DV_{\ov{0}}\oplus V_{\ov{1}}/DV_{\ov{1}}. \]
From eq. (\ref{df}) and (\ref{DistDer}) we get $(Du)_0=0$ and $u_0(Dv)=D(u_0v)$ for all $u,v\in V$ so that the product is well-defined. Equation (\ref{qs}) gives \[
u_0v+(-1)^{|u||v|}v_0u = 
-(-1)^{|u||v|}\sum_{1\leq k}(-1)^k\frac{D^k}{k!}(v_{k}u) \]
proving the graded skew-symmetry of the product. 
The graded Jacobi identity follows from the Jacobi identity on $V$ by taking 
$l=m=n=0$. \hspace*{\fill} $\Box$\\

The same equation shows that $V$ is a $V/DV$-module under the action $u.v=u_0v,\, u\in V/DV,v\in V$. The natural projection $V\rightarrow V/DV$ is a homomorphism of $V/DV$-modules.

\begin{pp1}
Let $V=V_{\ov{0}}\oplus V_{\ov{1}}$ and $W=W_{\ov{0}}\oplus W_{\ov{1}}$ be vertex superalgebras. Then 
\be 
(a\otimes b)_n (c\otimes d)=(-1)^{|b||c|}\sum_{k\in \mathbb Z}(a_kc)\otimes (b_{n-k-1}d)
\label{tp} \ee 
where $a,c\in V$ and $b,d\in W$, defines the structure of a vertex 
superalgebra on $V\otimes W$. The even part is $V_{\ov{0}}\otimes W_{\ov{0}}\oplus V_{\ov{1}}\otimes W_{\ov{1}}$ and the odd part $V_{\ov{0}}\otimes W_{\ov{1}}\oplus V_{\ov{1}}\otimes W_{\ov{0}}$.
The identity is given by $1\otimes 1$.
If $V$ and $W$ are simple so is $V\otimes W$.
\end{pp1}

An equivalent formulation of (\ref{tp}) is
\be Y(a\otimes b,z)=Y(a,z)\otimes Y(b,z)(-1)^{|b||.|}, \ee
since $(-1)^{|b||c|}\sum_{k\in \mathbb Z}(a_kc)\otimes (b_{n-k-1}d)$ is the coefficient of $z^{-n-1}$ in $Y(a,z)c\otimes (-1)^{|b||c|}Y(b,z)d$.

\section{Virasoro elements}

Let $V=V_{\ov{0}}\oplus V_{\ov{1}}$ be a vertex superalgebra. An element 
$\omega\in V_{\ov{0}}$ is called Virasoro element of central charge 
 $c\in {\mathbb R}$ if
\begin{enumerate}
\item The operators $L_m=\omega_{m+1}$ give a representation of the 
          Virasoro algebra
          \[ [L_m,L_n]=(m-n)L_{m+n}+\delta_{m+n,0}
                {\textstyle \frac{m^3-m}{12}}c. \] 
\item $V_{\ov{0}}$ and $V_{\ov{1}}$ can be decomposed into eigenspaces of 
          $L_0$
          \[ V_{\ov{0}}=\bigoplus_{n\in {\mathbb Z}}V_n \quad \mbox{and} \quad
             V_{\ov{1}}=\bigoplus_{n\in {\mathbb Z}+\frac{1}{2}}V_n\, , \]
          where $V_n=\{ v\in V | L_0v=nv \}$
\item $D=L_{-1}$
\end{enumerate}

Let $V$ be a vertex superalgebra with Virasoro element $\omega$. The operators $L_{-1},L_0$ and $L_1$ give a representation of the Lie algebra $sl_2\mathbb R$. If $v\in V_n$ then $L_1v\in V_{n-1}$ and $L_{-1}v\in V_{n+1}$. An element $v\in V$ is called quasiprimary if $L_1v=0$. The commutator relation of the $L_n$ implies
\bea
\omega_n\omega &=& \omega_n(\omega_{-1}1) \nonumber \\
               &=& [\omega_n,\omega_{-1}]1+\omega_{-1}(\omega_n1) \nonumber \\
     &=& (n+1)\omega_{n-2}1+\delta_{n,3}{\textstyle\frac{c}{2}}1
                                         +\omega_{-1}(\omega_n1),
\eea 
so that $L_0\omega=\omega_1\omega=2\omega$ and $\omega\in V_2\subset V_{\ov{0}}$ and $L_1\omega=\omega_2\omega=0$, i.e. $\omega$ is quasiprimary. 
From the commutator formula (\ref{comf}) and $\omega_m=L_{m-1}$ we get 
\be [L_{m-1},v_n]=\sum_{k\geq 0} {m \choose k}(L_{k-1}v)_{m+n-k}
     \; \mbox{ for all } v\in V. \label{iwp}
\ee
Eq. (\ref{df}) gives $[L_0,v_n]=-(n+1)v_n+pv_n$ for $v\in V_p$. It follows
\be 
\left( V_p \right)_n \left( V_q \right) \subset  V_{p+q-n-1} .
\ee
Let $P_n=\{v\in V| L_0v=nv,\, L_mv=0\, \mbox{ for all } m>0 \}$ for $n\in \frac{1}{2}{\mathbb Z}$. By (\ref{df})  
\be [L_m,v_n]=\{ (k-1)(m+1)-n \} v_{m+n} \; \mbox{ for all } v\in P_k. 
\label{vtrb} \ee
The representation of the Virasoro algebra on $V$ allows us to construct a subalgebra of $V/DV$.
\begin{pp1}
Let $V$ be a vertex superalgebra with Virasoro element. Then $P_1/DP_0$ is a Lie algebra.
\end{pp1}
{\it Proof:} Let $u,v\in P_1$ and $n\geq 0$. Then $[L_n,u_0]=0$ by (\ref{vtrb}) and $L_n(u_0v)=u_0(L_nv)=\delta_{n,0}u_0v$ by definition of $P_1$. This shows that $(P_1)_0(P_1)\subset P_1$. Next we prove $DV_{\ov{0}}\cap P_1=DP_0$. From $L_nL_{-1}v=(n+1)L_{n-1}v+L_{-1}L_nv$ for all $v\in V$ and $n\in {\mathbb Z}$ we get $DP_0\subset P_1$. Conversly let $v\in V_{\ov{0}}$ such that $Dv\in P_1$. Then $L_nL_{-1}v=0$ for $n\geq 1$ and $L_{n-1}v=-\frac{1}{n+1}L_{-1}L_nv$ for $n\geq 1$. There is an integer $k$ such that $L_mv=0$ for all $m\geq k$. It follows $L_mv=0$ for all $m\geq 0$, so that $v\in P_0$. \hfill $\Box$\\

Note that $P_1/DP_0$ is not a proper Lie superalgebra since $P_1$ is a subspace of $V_{\ov{0}}$. In bosonic string theory $P_1$ denotes the space of physical states, so that we will call $P_1/DP_0$ the Lie algebra of bosonic states. 

\begin{pp1} \label{tensorvir}
Let $V$ and $W$ be vertex superalgebras with Virasoro elements $\omega_V$ resp. $\omega_W$ of central charge $c$ resp. $\! d$. Then $V\otimes W$ is a vertex superalgebra with Virasoro element $\omega_V\otimes 1 + 1\otimes \omega_W$ of central charge $c+d$.
\end{pp1}      

\section{Neveu-Schwarz elements}

Let $V=V_{\ov{0}}\oplus V_{\ov{1}}$ be a vertex superalgebra. An element 
$\tau\in V_{\ov{1}}$ is called Neveu-Schwarz element of central charge 
 $c\in {\mathbb R}$ if 
\begin{enumerate}
\item $\omega={\frac{1}{2}}\tau_0\tau$ is a Virasoro element of
          central charge $c$.
\item The operators $L_n=\omega_{n+1}$ and $G_r=\tau_{r+\frac{1}{2}}$ 
          form a Neveu-Schwarz superalgebra
          \[ \renewcommand{\arraystretch}{1.5} \begin{array}{lcl}   
          {[}L_m,L_n{]} &=& (m-n)L_{m+n}+\delta_{m+n,0}\frac{m^3-m}{12}c \\
          {[}L_m,G_r{]} &=& (\frac{1}{2}m-r)G_{m+r} \\
          {[}G_r,G_s{]}_{+} &=& 2L_{r+s} + \frac{1}{3}(r^2-\frac{1}{4})
                                           \delta_{r+s,0}c 
             \end{array} \]
          
\end{enumerate}

Let $V$ be a vertex superalgebra with Neveu-Schwarz element $\tau$. The operators $L_0,L_{\pm 1}$ and $G_{\pm\frac{1}{2}}$ give a representation of the Lie superalgebra $osp(1,2)$. From the commutator relations of the $L_n$ and $G_r$ we get
\beann
\omega_n\tau &=& [\omega_n,\tau_{-1}]1+\tau_{-1}(\omega_n1) \\
             &=& ({\textstyle \frac{n}{2}}+1)\tau_{n-2}1+\tau_{-1}(\omega_n1)
\eeann
so that $L_0\tau=\frac{3}{2}\tau$ and $L_1\tau=0$, i.e. $\tau$ is quasiprimary. Furthermore
\beann
\tau_n\tau &=& [\tau_n,\tau_{-1}]_+1-\tau_{-1}(\tau_n1) \\
     &=& 2\omega_{n-1}1+(n^2-n)\delta_{n,2}
         {\textstyle\frac{c}{3}}1-\tau_{-1}(\tau_n1).
\eeann
As above we find
\be [G_{m-\frac{1}{2}},v_n]_{\mp}=\sum_{0\leq k} {m \choose k}
                                       (G_{k-\frac{1}{2}}v)_{m+n-k}
\ee
with $-$ for $|v|=0$ and $+$ for $|v|=1$. For $m=0$ we get 
$[G_{-\frac{1}{2}},v_n]_{\mp}=(G_{-\frac{1}{2}}v)_n$ or in terms of vertex operators  
\be [G_{-\frac{1}{2}},Y(v,z)]_{\mp}=Y(G_{-\frac{1}{2}}v,z). \ee
Define $\mbox{\sf P}_{n}=\{v\in V| L_0v=nv,\, L_mv=G_rv=0\, \mbox{ for all } m,r> 0 \}$ for $n\in \frac{1}{2}{\mathbb Z}$. As a consequence of the Neveu-Schwarz algebra $\mbox{\sf P}_{n}=\{v\in V| L_0v=nv,\, G_{\frac{1}{2}}v=G_{\frac{3}{2}}v=0 \}$. The following lemma describes the analogue of (\ref{vtrb}).
\begin{lp1} \label{comg}
Let $V$ be a vertex superalgebra with Neveu-Schwarz element. Then 
\be [G_{m-\frac{1}{2}},(G_{-\frac{1}{2}}v)_n]_{\mp}=\{2mk-(m+n)\}v_{m+n-1} \ee
for $v\in \mbox{\sf P}_{k}$ with $-$ for $k\in {\mathbb Z}$ and $+$ for 
$k\in {\mathbb Z}+\frac{1}{2}$.
\end{lp1} 
{\it Proof:}
Let $v\in \mbox{\sf P}_{k}$. Then $G_{\frac{1}{2}}G_{-\frac{1}{2}}v=2L_0v-G_{-\frac{1}{2}}G_{\frac{1}{2}}v=2kv$ and $G_rG_{-\frac{1}{2}}v=2L_{r-\frac{1}{2}}v-G_{-\frac{1}{2}}G_rv=0$ for $r>\frac{1}{2}$, so that
\beann
\lefteqn{ [G_{m-\frac{1}{2}},(G_{-\frac{1}{2}}v)_n]_{\mp} } \\
&=& \sum_{k\geq 0} {m \choose k} (G_{k-\frac{1}{2}}G_{-\frac{1}{2}}v)_{m+n-k}\\
&=& (G_{-\frac{1}{2}}G_{-\frac{1}{2}}v)_{m+n}
          +m(G_{\frac{1}{2}}G_{-\frac{1}{2}}v)_{m+n-1} \\
&=& (L_{-1}v)_{m+n}+2mkv_{m+n-1} \\
&=& (Dv)_{m+n}+(m+n)v_{m+n-1}+\{2mk-(m+n)\}v_{m+n-1}\\
&=& \{2mk-(m+n)\}v_{m+n-1}.
\eeann
\hspace*{\fill} $\Box$\\
We now construct a subalgebra of $P_1/DP_0$ with the help of the Neveu-Schwarz
algebra.
\begin{pp1}
Let $V$ be a vertex superalgebra with Neveu-Schwarz element. Then \\
$G_{-\frac{1}{2}}\mbox{\sf P}_{\frac{1}{2}}/D\mbox{\sf P}_0$
is a Lie algebra.
\end{pp1}
{\it Proof:} Let $u,v\in \mbox{\sf P}_{\frac{1}{2}}$. Then
\beann
(G_{-\frac{1}{2}}u)_0(G_{-\frac{1}{2}}v) 
&=& [G_{-\frac{1}{2}},u_0]_+G_{-\frac{1}{2}}v \\
&=& G_{-\frac{1}{2}}(u_0G_{-\frac{1}{2}}v)+
            u_0(G_{-\frac{1}{2}}G_{-\frac{1}{2}}v) \\
&=& G_{-\frac{1}{2}}(u_0G_{-\frac{1}{2}}v)+u_0L_{-1}v \\
&=& G_{-\frac{1}{2}}(u_0G_{-\frac{1}{2}}v)+L_{-1}u_0v \\
&=& G_{-\frac{1}{2}}(u_0G_{-\frac{1}{2}}v+G_{-\frac{1}{2}}u_0v) \\
&=& G_{-\frac{1}{2}}([G_{-\frac{1}{2}},u_0]_+v) \\
&=& G_{-\frac{1}{2}}((G_{-\frac{1}{2}}u)_0v),
\eeann
since $[u_0,L_{-1}]=0$ by (\ref{vtrb}). From $L_0(G_{-\frac{1}{2}}u)=[L_0,G_{-\frac{1}{2}}]u+G_{-\frac{1}{2}}L_0u=G_{-\frac{1}{2}}u$ follows $(G_{-\frac{1}{2}}u)_0v\in V_{1+\frac{1}{2}-1}=V_{\frac{1}{2}}$. Now by Lemma (\ref{comg}) $G_r$ anticommutes with $(G_{-\frac{1}{2}}u)_0$, so that $G_r((G_{-\frac{1}{2}}u)_0v)=-(G_{-\frac{1}{2}}u)_0(G_rv)=0$ for all $r>0$. Hence $(G_{-\frac{1}{2}}u)_0v$ is in $\mbox{\sf P}_{\frac{1}{2}}$ and $(G_{-\frac{1}{2}}\mbox{\sf P}_{\frac{1}{2}})_0(G_{-\frac{1}{2}}\mbox{\sf P}_{\frac{1}{2}} )\subset (G_{-\frac{1}{2}}\mbox{\sf P}_{\frac{1}{2}})$. 
Let $v\in \mbox{\sf P}_{\frac{1}{2}}$. Then $L_0(G_{-\frac{1}{2}}v)=
G_{-\frac{1}{2}}v$ and $L_n(G_{-\frac{1}{2}}v)=[L_n,G_{-\frac{1}{2}}]v+G_{-\frac{1}{2}}L_nv=\frac{1}{2}(n+1)G_{n-\frac{1}{2}}v+G_{-\frac{1}{2}}L_nv=0$ for $n>0$ show that 
\be G_{-\frac{1}{2}}\mbox{\sf P}_{\frac{1}{2}} \subset P_1\, . \ee
It remains to show that 
$DP_0\cap G_{-\frac{1}{2}}\mbox{\sf P}_{\frac{1}{2}} =D\mbox{\sf P}_0$.
Let $v\in \mbox{\sf P}_0$. Then $L_0G_{-\frac{1}{2}}v=[L_0,G_{-\frac{1}{2}}]v+G_{-\frac{1}{2}}L_0v=\frac{1}{2}G_{-\frac{1}{2}}v$ and $G_rG_{-\frac{1}{2}}v=2L_{r-\frac{1}{2}}v-G_{-\frac{1}{2}}G_rv=0$ for $r>0$, so that $G_{-\frac{1}{2}} \mbox{\sf P}_0 \subset \mbox{\sf P}_{\frac{1}{2}}$. Since $\mbox{\sf P}_0$ is a subspace of $P_0$ we have 
\[
G_{-\frac{1}{2}} \mbox{\sf P}_0 \subset \left( \mbox{\sf P}_{\frac{1}{2}} 
        \cap G_{-\frac{1}{2}}P_0 \right). \]
Applying $G_{-\frac{1}{2}}$ to both sides gives
\[
G_{-\frac{1}{2}} G_{-\frac{1}{2}} \mbox{\sf P}_0 \subset 
G_{-\frac{1}{2}} \left( \mbox{\sf P}_{\frac{1}{2}} 
                          \cap G_{-\frac{1}{2}}P_0 \right) \subset
\left( G_{-\frac{1}{2}}\mbox{\sf P}_{\frac{1}{2}}\cap
       G_{-\frac{1}{2}}G_{-\frac{1}{2}}P_0 \right). \]
$D=G_{-\frac{1}{2}}^{\;\, 2}$ implies $D\mbox{\sf P}_0\subset 
(G_{-\frac{1}{2}}\mbox{\sf P}_{\frac{1}{2}}\cap DP_0)$. 
Now let $u\in P_0$ such that there is a $v\in \mbox{\sf P}_{\frac{1}{2}}$ with
$Du=G_{-\frac{1}{2}}v$. Then $G_rDu=G_rG_{-\frac{1}{2}}v=-G_{-\frac{1}{2}}G_rv+
2L_{r-\frac{1}{2}}v=0$ for $r>\frac{1}{2}$. From $[D,G_r]=(-\frac{1}{2}-r)G_{r-1}$ for all $r$, we get $G_{r-1}u=-\frac{1}{r+\frac{1}{2}}DG_ru$ for all $r>\frac{1}{2}$. Since there is an $s$ such that $G_ru=0$ for all $r>s$, this formula implies $G_ru=0$ for $r>0$. Hence $u\in \mbox{\sf P}_0$ and $(DP_0\cap G_{-\frac{1}{2}}\mbox{\sf P}_{\frac{1}{2}})\subset D\mbox{\sf P}_0$. 
\hspace*{\fill} $\Box$\\

$\mbox{\sf P}_{\frac{1}{2}}$ can be interpreted 
as the space of physical states of the Neveu-Schwarz superstring. We will call
$G_{-\frac{1}{2}}\mbox{\sf P}_{\frac{1}{2}}/D\mbox{\sf P}_0$ Lie algebra of 
Neveu-Schwarz states. 

Note that $G_{-\frac{1}{2}}$ is injective on $\mbox{\sf P}_{\frac{1}{2}}$ for if $v\in \mbox{\sf P}_{\frac{1}{2}}$ such that $G_{-\frac{1}{2}}v=0$ then
$0=G_{\frac{1}{2}}G_{-\frac{1}{2}}v=[G_{\frac{1}{2}},G_{-\frac{1}{2}}]_+v
=2L_0v=v$.

The following analogue of Proposition (\ref{tensorvir}) is easy to prove.
\begin{pp1}
Let $V$ and $W$ be vertex superalgebras with Neveu-Schwarz elements $\tau_V$ resp. $\tau_W$ of central charge $c$ resp. $d$. Then $V\otimes W$ is a vertex superalgebra with Neveu-Schwarz element $\tau_V\otimes 1 + 1\otimes \tau_W$ of central charge $c+d$.
\end{pp1}    

\section{Invariant bilinear forms}\label{inff}

Let $V$ be a vertex superalgebra with Virasoro element. Suppose that $L_1$ acts locally nilpotent on $V$. Fix a nonzero constant $\lambda$ and define the opposite vertex operator of $v\in V$ by
\be 
Y^*(v,z)=Y(e^{-\lambda^{-2}zL_1}(-\lambda^{-1}z)^{-2L_0}v,-\lambda^2z^{-1})
\ee 
and the components $v_n^*$ of $v$ by
\be Y^*(v,z)=\sum_{n\in \mathbb Z}v_n^* z^{-n-1}. \ee
If $v \in V_p$ then 
\be v_n^*=\lambda^{-2p} (-\lambda^{2})^{n+1}
          \sum_{m\geq 0} \left( \frac{L_1^m}{m!}v \right)_{2p-n-m-2} 
\label{dgt} \ee
is a well-defined endomorphism of $V$ and
\be v_n^*w \in V_{q-p+n+1} \ee
for $w \in V_q$.
For a quasiprimary element $v\in V_p$ equation (\ref{dgt}) reduces to
\be v_n^*=\lambda^{-2p} (-\lambda^{2})^{n+1}v_{2p-n-2}, \ee
e.g. $\omega_n^*=\lambda^{-4}(-\lambda^{2})^{n+1}\omega_{2-n}$. If we define 
$L_n^*=\omega_{n+1}^*$ then $L_n^*= (-\lambda^{2})^n L_{-n}$. 
The opposite vertex operators satisfy the following conjugation formula 
(cf. \cite{K2})
\be e^{\lambda L_{-1}}e^{\lambda^{-1} L_1}e^{\lambda L_{-1}} Y(v,z)
    e^{-\lambda L_{-1}}e^{-\lambda^{-1} L_1}e^{-\lambda L_{-1}}
   =Y^*(v,z). \label{kacconjf} \ee
It implies that $v_n^*w=0$ for $n$ big enough.
Further properties of opposite vertex operators are summarized in the 
following 
\begin{pp1}
Let $V$ be a vertex superalgebra with Virasoro element such that $L_1$ acts locally nilpotent. Then for $u,v$ in $V$ and $l,m,n$ in $\mathbb Z$ 
\bea 
1_n^*       &=& (-\lambda^2)^{n+1}\delta_{n+1,0} \label{eins*} \\
(Du)_n^*    &=& -nu_{n-1}^* \label{deropp} \\
\lefteqn{\sum_{0\leq k} {m \choose k}(v_{l+k}u)_{m+n-k}^*} \label{oji} \\
            &=&    \sum_{0\leq k} {l \choose k}(-1)^k \left\{
                   (-1)^{|u||v|}u_{n+k}^*v_{l+m-k}^*-(-1)^lv_{m+k}^*u_{l+n-k}^*
                    \right\} \nonumber
\eea
\end{pp1}
{\it Proof:} The equation $1_n = \delta_{n+1,0}$ implies (\ref{eins*}). 
Since the operators $L_{-1},L_0$ and $L_1$ give a representation of $sl_2$ they
satisfy (cf. \cite{FHL})
\[ L_{-1}L_1^n=L_1^nL_{-1}-2nL_1^{n-1}L_0+n(n-1)L_1^{n-1}. \]
Inserting this into (\ref{dgt}) yields the second equation. 
The Jacobi identity on $V$ gives
\[ 
\renewcommand{\arraystretch}{1.8}
\begin{array}{ll} 
\multicolumn{2}{l}
{ \D
\left( \frac{\lambda^2z_0}{z_1z_2} \right)^{-1}
   \delta\left( \frac{-\lambda^2/z_1+\lambda^2/z_2}
                     { \lambda^2z_0/z_1z_2        }  \right)
   Y\big(e^{-\lambda^{-2}z_1L_1}(-\lambda^{-1}z_1)^{-2L_0}u,-\lambda^2/z_1
    \big)
} \\
\multicolumn{2}{l}
{ \D
\qquad \qquad 
   Y\big(e^{-\lambda^{-2}z_2L_1}(-\lambda^{-1}z_2)^{-2L_0}v,-\lambda^2/z_2
   \big)  
} \\
\multicolumn{2}{l}
{ \D
-(-1)^{|u||v|} \left( \frac{\lambda^2z_0}{z_1z_2} \right)^{-1}
   \delta\left( \frac{-\lambda^2/z_2+\lambda^2/z_1 }
                     {-\lambda^2z_0/z_1z_2          } \right) 
} \\
\multicolumn{2}{l}
{ \D
Y\big(e^{-\lambda^{-2}z_2L_1}(-\lambda^{-1}z_2)^{-2L_0}v,-\lambda^2/z_2
    \big) 
   Y\big(e^{-\lambda^{-2}z_1L_1}(-\lambda^{-1}z_1)^{-2L_0}u,-\lambda^2/z_1
    \big)
} \\
\D = & \D \left( -\frac{\lambda^2}{z_2} \right)^{-1}
    \delta\left( \frac{-\lambda^2/z_1-\lambda^2z_0/z_1z_2}
                      {-\lambda^2/z_2} \right) \\
     & \D Y\Big( Y\big(e^{-\lambda^{-2}z_1L_1}(-\lambda^{-1}z_1)^{-2L_0}u, 
             \lambda^2z_0/z_1z_2\big)  \\
     & \D \qquad\qquad\quad
         e^{-\lambda^{-2}z_2L_1}(-\lambda^{-1}z_2)^{-2L_0}v,
                          -\lambda^2/z_2
    \Big) .
\end{array} \]
Multiplying both sides with $-\lambda^2/z_1z_2$ we get
\bea
&& (-1)^{|u||v|} z_0^{-1}\delta\left(\frac{z_1-z_2}{z_0}\right)
            Y^*(v,z_2)Y^*(u,z_1) \label{jf} \\
&& -z_0^{-1}\delta\left(\frac{z_2-z_1}{-z_0}\right)
            Y^*(u,z_1)Y^*(v,z_2) \nonumber \\
&=& z_1^{-1}\delta\left(\frac{z_2+z_0}{z_1}\right) 
    Y\Big( Y\big(e^{-\lambda^{-2}z_1L_1}(-\lambda^{-1}z_1)^{-2L_0}u, 
                      \lambda^2z_0/z_1z_2 \big) \nonumber \\ 
&&  \qquad\qquad\qquad\qquad\qquad
    e^{-\lambda^{-2}z_2L_1}(-\lambda^{-1}z_2)^{-2L_0}v,-\lambda^2/z_2
     \Big) \nonumber .
\eea
The equations
\beann
z^{2L_0}Y(v,z_0)z^{-2L_0} &=& Y(z^{2L_0}v,z^2z_0) \\
e^{zL_1}Y(v,z_0)z^{-zL_1} &=& Y(e^{z(1-zz_0)L_1}(1-zz_0)^{-2L_0}v,z/1-zz_0) 
\eeann
(cf. \cite{FHL} and \cite{K2}) imply 
\beann
\lefteqn{e^{-\lambda^{-2}z_2L_1}(-\lambda^{-1}z_2)^{-2L_0}Y(u,z_0)
                         (-\lambda^{-1}z_2) ^{2L_0}e^{\lambda^{-2}z_2L_1}}\\
&=& Y\Big(e^{-\lambda^{-2}(z_2+z_0)L_1}(-\lambda^{-1}(z_2+z_0))^{-2L_0}u,
           \lambda^2z_0/(z_2+z_0)z_2 \Big) 
\eeann
so that 
\beann
\lefteqn{\hspace{-3.5cm}
 Y\Big(e^{-\lambda^{-2}(z_2+z_0)L_1}(-\lambda^{-1}(z_2+z_0))^{-2L_0}u,
 \lambda^2z_0/(z_2+z_0)z_2 \Big)
               e^{-\lambda^{-2}z_2L_1}(-\lambda^{-1}z_2)^{-2L_0}} \\
&=& e^{-\lambda^{-2}z_2L_1}(-\lambda^{-1}z_2)^{-2L_0}Y(u,z_0) 
\eeann
and 
\beann
& & z_1^{-1}\delta\left( \frac{z_2+z_0}{z_1}\right) 
    Y\Big( Y \big( e^{-\lambda^{-2}z_1L_1}(-\lambda^{-1}z_1)^{-2L_0}u, 
                  \lambda^2z_0/z_1z_2 \big) \\  
& & \qquad\qquad
     e^{-\lambda^{-2}z_2L_1}(-\lambda^{-1}z_2)^{-2L_0}v,-\lambda^2/z_2
    \Big) \\
&=& z_1^{-1}\delta\left(\frac{z_2+z_0}{z_1}\right) \\
& & Y\Big( 
    Y\big( e^{-\lambda^{-2}(z_2+z_0)L_1}(-\lambda^{-1}(z_2+z_0))^{-2L_0}u, 
           \lambda^2z_0/(z_2+z_0)z_2 \big) \\
& & \qquad\qquad
    e^{-\lambda^{-2}z_2L_1}(-\lambda^{-1}z_2)^{-2L_0}v,
                                          -\lambda^2/z_2\Big)\\
&=& z_1^{-1}\delta\left(\frac{z_2+z_0}{z_1}\right)
     Y\big( e^{-\lambda^{-2}z_2L_1}(-\lambda^{-1}z_2)^{-2L_0}Y(u,z_0)v,
                                                   -\lambda^2/z_2\big)\\
&=& z_2^{-1}\delta\left(\frac{z_1-z_0}{z_2}\right)Y^*(Y(u,z_0)v,z_2)
\eeann
where we have used $z_1^{-1}\delta(\frac{z_2+z_0}{z_1})=z_2^{-1}\delta(\frac{z_1-z_0}{z_2})$. Putting this into (\ref{jf}) gives
\beann
&& (-1)^{|u||v|} z_0^{-1}\delta\left(\frac{z_1-z_2}{z_0}\right)
                          Y^*(v,z_2)Y^*(u,z_1) \\
&& -z_0^{-1}\delta\left( \frac{z_2-z_1}{-z_0}\right) Y^*(u,z_1)Y^*(v,z_2) \\
&=& z_2^{-1}\delta\left( \frac{z_1-z_0}{z_2}\right)Y^*(Y(u,z_0)v,z_2). 
\eeann 
Multiplying this equation with $z_0^lz_1^mz_2^n$ and evaluating 
$Res_{z_2}Res_{z_1}Res_{z_0}$ of this expression gives the last equation of the proposition with $u$ and $v$ interchanged. \hspace*{\fill} $\Box$\\

Putting $l=0$ in (\ref{oji}) gives the commutator formula
\be [v_m^*,u_n^*]_{\mp} = -\sum_{0\leq k}{m \choose k}(v_ku)_{m+n-k}^* . \ee
Taking $v=\omega$ and using (\ref{deropp}) yields
\be [L_1,u_n^*]=-\lambda^2 nu_{n-1}^* \label{dumdidel}. \ee
As a consequence we get $u_n^*1=0$ for $n\geq 0$. 
Putting $m=0$ in (\ref{oji}) gives the associativity formula
\be
(v_lu)_n^* = \sum_{0\leq k} {l \choose k}(-1)^k \left\{
                   (-1)^{|u||v|}u_{n+k}^*v_{l-k}^*-(-1)^lv_{k}^*u_{l+n-k}^*
                    \right\}  \label{oaf}
\ee
A bilinear form $(\, , \, )$ on $V$ is called invariant if
\be (u_nv,w)=(-1)^{|u||v|}(v,u_n^*w) \label{surfen} \ee
for $n\in \mathbb Z$.
Note that the invariance condition implies $(L_nV_n,1)=0$ for $n\leq 1$.
\begin{lp1}\label{isst}
Let $m\geq 0$. Then for $p\in \frac{1}{2}{\mathbb Z}$ and $u,v\in V_p$
\be (L_1^mu)_{2p-m-1}v-(L_1^mv)_{2p-m-1}u \in L_1V_1+L_{-1}V_{-1} 
\label{ffi}. \ee
\end{lp1}
{\it Proof:} By induction. Note that $(-1)^{2p}(-1)^{|u||v|}=1$ so that formula (\ref{qs}) gives
\[ u_{2p-1}v-v_{2p-1}u = 
\sum_{j\geq 1} (-1)^j \frac{L_{-1}^j}{j!} v_{2p-1+j}u \]
which is in $L_{-1}V_{-1}$. This proves the case $m=0$. From (\ref{iwp}) we get
\be [L_1,u_n]=\{2p-(n+2)\}u_{n+1}+(L_1u)_n, \label{wip} \ee
so that 
\beann
(L_1u)_{2p-2}v-(L_1v)_{2p-2}u &=& [L_1,u_{2p-2}]v-[L_1,v_{2p-2}]u \\
&=& L_1u_{2p-2}v-L_1v_{2p-2}u-u_{2p-2}L_1v+v_{2p-2}L_1u.
\eeann
Using again (\ref{qs}) we find 
$-u_{2p-2}L_1v+v_{2p-2}L_1u=-u_{2p-2}L_1v-(L_1u)_{2p-2}v$ modulo some element in $L_{-1}V_{-1}$. But $u_{2p-2}L_1v+(L_1u)_{2p-2}v=L_1u_{2p-2}v$. This shows
that $(L_1u)_{2p-2}v-(L_1v)_{2p-2}u\in L_1V_1+L_{-1}V_{-1}$ and (\ref{ffi}) is true for $m=1$.
Now suppose that (\ref{ffi}) is true for $m-1$. Then by (\ref{wip})
\beann
\lefteqn{(L_1^mu)_{2p-m-1}v-(L_1^mv)_{2p-m-1}u =} \\
& & (m-1)\left\{ (L_1^{m-1}u)_{2p-m}v-(L_1^{m-1}v)_{2p-m}u \right\} \\
& & +[L_1,(L_1^{m-1}u)_{2p-m-1}]v-[L_1,(L_1^{m-1}v)_{2p-m-1}]u.
\eeann
The first term is in $L_1V_1+L_{-1}V_{-1}$ by the induction hypothesis. 
Thus we only need to consider the term
$-(L_1^{m-1}u)_{2p-m-1}L_1v+(L_1^{m-1}v)_{2p-m-1}L_1u$. $L_1u$ and $L_1v$ are in $V_{p-1}$. Applying the induction hypothesis with $m-2$ on these elements shows that this term is also in $L_1V_1+L_{-1}V_{-1}$. \hspace*{\fill} $\Box$\\

Now we can prove the following
\begin{pp1}
Let $V$ be a vertex superalgebra with Virasoro element such that $L_1$ acts locally nilpotent and let $(\, , \, )$ be an invariant form on $V$. Then
\begin{enumerate}
\item $(V_p,V_q)=0$ for $p\neq q$
\item $(u,v)=(v,u)$ for $u,v\in V$
\end{enumerate}
\end{pp1}
{\it Proof:} Let $u\in V_p$ and $v\in V_q$. Then $0=(L_0u,v)-(u,L_0v)=
(p-q)(u,v)$ since $L_0^*=L_0$. It is sufficient to prove the symmetry 
for $u,v\in V_p$. By Lemma \ref{isst} 
\beann
u_{-1}^*v-v_{-1}^*u &=&
\lambda^{-2p} \bigg\{ 
\sum_{m\geq 0} \bigg( \frac{L_1^m}{m!}u \bigg)_{2p-m-1}v
-\sum_{m\geq 0} \bigg( \frac{L_1^m}{m!}v \bigg)_{2p-m-1}u \bigg\}
\eeann
is an element of $L_1V_1+L_{-1}V_{-1}$ so that 
$(u,v)-(v,u)=(1,u_{-1}^*v-v_{-1}^*u)=0$.
\hspace*{\fill} $\Box$\\

Define a linear functional $f:V_0\rightarrow {\mathbb R}$ by $f(u)=(1,u)=(u,1)$ for $u\in V_0$. Let $M$ be the vector space generated by the elements $u_n^*v$ with $u\in V_p,\, v\in V_q, \, q-p+n+1=0$ and $n\geq 0$. Then $0=(u_n1,v)=(1,u_n^*v)=f(u_n^*v)$ shows that M is in the kernel of $f$. We will now show that
\be L_1V_1=L_1V_1+L_{-1}V_{-1}=M.\ee
To do so we will establish 
$L_1V_1\subset L_1V_1+L_{-1}V_{-1}\subset M\subset L_1V_1$. The first inclusion is clear. The second follows from $L_nV_n=\omega_{n+1}V_n=\omega_{1-n}^*V_n\subset M$ for $1-n\geq 0$ resp. $n\leq 1$. Let $u\in V_p,\, v\in V_q, \, q-p+n+1=0$ and $n\geq 0$. Then by (\ref{dumdidel})
\beann
u_n^*v &=& -\frac{1}{\lambda^2 (n+1)}[L_1,u_{n+1}^*]v \\
       &=& -\frac{1}{\lambda^2 (n+1)}(L_1u_{n+1}^*v-u_{n+1}^*L_1v).
\eeann
Applying the same argument to $u_{n+1}^*L_1v, u_{n+2}^*L_1^2v, \ldots$ we
find that $u_n^*v$ is in $L_1V_1$ since $L_1^kv=0$ for $k$ sufficiently large.

\begin{thp1}
Let $V$ be a vertex superalgebra with Virasoro element such that $L_1$ acts locally nilpotent. Then the space of invariant bilinear forms on $V$ is naturally
isomorphic to the dual of $V_0/L_1V_1$.
\end{thp1}
{\it Proof:} We have already shown that an invariant form determines a functional on $V_0$ vanishing on $L_1V_1$. Conversly let $f$ be in the dual of $V_0/L_1V_1$. Let $\pi$ be the natural projection of $V$ on $V_0/L_1V_1$. Then $\pi(u_{-1}^*v)=0$ for $u\in V_p,\, v\in V_q$ and $p\neq q$. Define
\[ (u,v)=(f\circ \pi) (u_{-1}^*v) \] for $u,v\in V$. Then $(\, , \, )$ is an invariant form on $V$ as we will see. Let $u\in V_p,\, v\in V_q$ and $w\in V_r$.
Then $(u_nv,w)$ and $(v,u_n^*w)$ are only nonzero if $p+q-r=n+1$. Suppose that this relation is satisfied. By the associativity formula (\ref{oaf})
\pagebreak
\beann
\lefteqn{ (u_nv)_{-1}^*-(-1)^{|u||v|}v_{-1}^*u_{n}^*} \\
&=& \sum_{0\leq k} {n \choose k}(-1)^k \left\{
                   (-1)^{|u||v|}v_{k-1}^*u_{n-k}^*-(-1)^nu_{k}^*v_{n-k-1}^*
                    \right\} \\
& & -(-1)^{|u||v|}v_{-1}^*u_{n}^*\\
&=& -(-1)^nu_0^*v_{n-1}^* \\
& & +\sum_{1\leq k} {n \choose k}(-1)^k \left\{
                   (-1)^{|u||v|}v_{k-1}^*u_{n-k}^*-(-1)^nu_{k}^*v_{n-k-1}^*
                    \right\},
\eeann
so that $\{ (u_nv)_{-1}^*-(-1)^{|u||v|}v_{-1}^*u_{n}^* \}w$ is in $M=L_1V_1$ and
\beann 
\lefteqn{ (u_nv,w)-(-1)^{|u||v|}(v,u_n^*w) } \\
 &=& (1,\{(u_nv)_{-1}^*-(-1)^{|u||v|}v_{-1}^*u_{n}^* \}w) \\
 &=& 0\, .
\eeann 
\hspace*{\fill} $\Box$\\

Two remarks should be made on the constant $\lambda$. The above results remain true over the complex numbers. If $V$ is a complex vertex superalgebra with Virasoro element such that $L_1$ is locally nilpotent then the choice $\lambda =i$
gives nice formulas, e.g. (\ref{dgt}) simplifies to 
\be 
v_n^*=i^{-2p} \sum_{m\geq 0} \left( \frac{L_1^m}{m!}v \right)_{2p-n-m-2} 
\label{vac} \ee  
for $v\in V_p$. If $V$ is a real vertex algebra, i.e. a real vertex superalgebra with $V_{\ov{1}}=0$, with Virasoro element such that $L_1$ is locally nilpotent then (\ref{vac}) still makes sense and gives the usual definition of opposite vertex operators. Similar arguments hold in the case that we are only interested in invariant forms on the even part of a vertex superalgebra.

\section{Bosonic construction} \label{bosc}

In this section we construct the vertex superalgebra corresponding to an integral lattice. The momenta of a bosonic string moving on a torus ("compactified bosonic string") lie on a lattice, so that this vertex algebra can be regarded as the Fock space of such a string.

Let $L$ be an integral lattice of rank $d$ with nondegenerate bilinear form $\la \, ,\, \ra$. We have the coset decomposition  
\[ L=L_0 \cup L_1 \] with 
\[ L_i=\{ \alpha \in L | \la \alpha ,\alpha \ra \in 2{\mathbb Z}+i \} 
             \quad \mbox{for}\; i=0,1.  \]
We put $h=L{\otimes}_{\mathbb Z}{\mathbb R}$ and extend the symmetric form of 
$L$ to $h$. Define the infinite dimensional Heisenberg algebra 
\be \hat{h}=h\otimes {\mathbb R}[t,t^{-1}]\oplus {\mathbb R}c \ee
with products
\be [h_1(m),h_2(n)]=m\delta_{m+n,0} \la h_1,h_2 \ra c, \quad [h_1(m),c]=0 \ee
for $h_1,h_2\in h,\, m,n\in {\mathbb Z}$. Then 
\be \hat{h}^-=h\otimes t^{-1}{\mathbb R}[t^{-1}] \ee
is an abelian subalgebra of $\hat{h}$, and $S(\hat{h}^-)$ is the symmetric algebra of polynomials in $\hat{h}^-$.  
  
Let ${\mathbb R}[L]$ be the group algebra of $L$ with basis $\{ e^{\alpha} 
|\: \alpha \in L \}$ and products $e^{\alpha}e^{\beta}=e^{\alpha + \beta}$. 
The vector space 
\be V=S(\hat{h}^-)\otimes {\mathbb R}[L] \ee
decomposes as
\be V=V_{\ov{0}}\oplus V_{\ov{1}} \ee
where 
\be V_{\ov{\imath}}=S(\hat{h}^-)\otimes {\mathbb R}[L_i]. \ee

$\hat{h}$ acts on $V$ as follows. For $0<m\in{\mathbb Z},\, k\in h,\, u\in S(\hat{h}^-),\, \alpha\in L$ let 
\bea
 k(m)\cdot (u\otimes e^{\alpha}) &=& m(\partial_{k(-m)}u)\otimes e^{\alpha} \\ k(-m)\cdot (u\otimes e^{\alpha}) &=& (k(-m)u)\otimes e^{\alpha} \\
 k(0)\cdot (u\otimes e^{\alpha}) &=& \la k,\alpha \ra u\otimes e^{\alpha} \\
 c\cdot (u\otimes e^{\alpha}) &=& u\otimes e^{\alpha}.
\eea

Choose a bimultiplicative map $\varepsilon : L\times L \rightarrow \{\pm 1\}$ with the properties (cf. \cite{K2} p. 108)
\bea
\varepsilon(\alpha,\beta)\varepsilon(\beta,\alpha)^{-1} &=&
(-1)^{\la \alpha, \beta \ra + \la \alpha, \alpha \ra \la \beta, \beta \ra }\\
\varepsilon(\alpha,\alpha) &=& (-1)^{\frac{1}{2}(\la \alpha, \alpha\ra
                      +\la \alpha, \alpha\ra^2) }\, .
\eea
Note that $\varepsilon$ is a 2-cocylce and satisfies
\be \varepsilon(0,0)=\varepsilon(\alpha,0)=\varepsilon(0,\alpha)=1 \, . 
\label{oa} \ee
This implies
\[ 1=\varepsilon(\alpha,0)=\varepsilon(\alpha,\alpha-\alpha)=
     \varepsilon(\alpha,\alpha)\varepsilon(\alpha,-\alpha) \]
and 
\be \varepsilon(\alpha,\alpha)=\varepsilon(\alpha,-\alpha)\, . \ee
Define a linear operator $\varepsilon_{\alpha} : V\rightarrow V$ for $\alpha \in L$ by 
\be \varepsilon_{\alpha}(u\otimes e^{\beta})=\varepsilon (\alpha,\beta)u\otimes e^{\beta}. \ee
For $v=1\otimes e^{\alpha}$ put
\be
 Y(v,z) = \exp \left(\sum_{m\geq 1}\alpha (-m)\frac{z^m}{m} \right) e^{\alpha}z^{\alpha(0)} \exp \left(-\sum_{m\geq 1}\alpha (m)\frac{z^{-m}}{m} \right) 
\varepsilon_{\alpha}\; . \label{iwsurf}
\ee
The first and the last exponentials act on $S(\hat{h}^-)$. The middle operators act on ${\mathbb R}[L]$ by
\be e^{\alpha}z^{\alpha(0)}\cdot e^{\beta}=z^{\la \alpha, \beta \ra} e^{\alpha+\beta}. \ee 
Expression (\ref{iwsurf}) is usually abbreviated 
$Y(1\otimes e^{\alpha},z)=\;: \exp (i\alpha X(z)):\,\varepsilon_{\alpha}.$
By (\ref{oa}) we have 
\be Y(1\otimes e^0,z)=1. \ee
For $v=k(-n-1)\otimes e^0$ and $n\geq0$ we define 
\be Y(v,z)=\frac{1}{n!}\left(\dz \right)^n \left( \frac{k(z)}{z} \right) \ee
with
\be k(z)=\sum_{n\in {\mathbb Z}}k(n)z^{-n} . \ee
  
We now define the bosonic normal ordering for any product of Heisenberg generators by
\be : h_1(n_1)\cdots h_r(n_r) := h_{\sigma 1}(n_{\sigma 1})\cdots 
h_{\sigma r}(n_{\sigma r}), \ee
where $\sigma \in S_r$ is any permutation such that $n_{\sigma 1}\leq \ldots \leq n_{\sigma r}$. This is equivalent to putting all $h(n)$ with $n<0$ ("creation operators") to the left of those with $n\geq 0$ ("annihilation operators"). The operators $e^{\alpha}$ commute with $h(n)$ for $n\not= 0$. Define the bosonic normal ordering of any product involving operators $e^{\alpha}, z^{\beta(0)}$ and $h(n)$ to be the rearranged product with the $h(n)$ normally ordered as above and $e^{\alpha}$ to the left of all $z^{\beta(0)}$ and all $h(0)$. 

For an element of the form
\be v=h_1(-n_1)\cdots h_r(-n_r)\otimes e^{\alpha} \ee define
\be Y(v,z)=:Y(h_1(-n_1)\otimes e^0,z)\cdots Y(h_r(-n_r)\otimes e^0,z)Y(1\otimes
              e^{\alpha},z): \label{sogehts} \ee 
and extend this definition linearly. 
This gives us a well-defined linear map
\beann 
Y\; :\; V &\rightarrow & End(V)[\![z,z^{-1}]\!] \\
        v &\mapsto     & Y(v,z)=\sum_{n\in {\mathbb Z}}v_nz^{-n-1}, 
                                 \quad v_n\in End(V)
\eeann
\begin{pp1} 
Y provides $V=V_{\ov{0}}\oplus V_{\ov{1}}$ with the structure of a vertex superalgebra. The identity is given by $1\otimes e^0$. V is simple.
\end{pp1} 

As we will see $V$ even contains a Virasoro element. 

Let $\{h^1,\ldots,h^d\}$ be a basis of $h$ and $\{h_1,\ldots,h_d\}$ be the corresponding dual basis of $h$ defined by 
\[ \la h^i,h_j \ra = \delta_{ij} \quad i,j=1,\ldots,d. \]
Then the element 
\be \omega={\textstyle \frac{1}{2}}\sum_{i=1}^{d}h^i(-1)h_i(-1) \otimes e^0
                \in V_{\ov{0}} \ee
is independent of the choice of basis of $h$.
Using (\ref{sogehts}) we get
\beann 
Y(\omega,z) &=& {\textstyle \frac{1}{2}}\sum_{m,n\in {\mathbb Z}}
                  \, \sum_{i=1}^d :h^i(m)h_i(n):z^{-(m+n)-2} \\
            &=& {\textstyle \frac{1}{2}}\sum_{k\in {\mathbb Z}}\,
             \sum_{m\in {\mathbb Z}}\,\sum_{i=1}^d :h^i(m)h_i(k-m):z^{-k-2} \\
            &=& \sum_{k\in {\mathbb Z}}L_k z^{-k-2},
\eeann
where we have defined
\be L_k=\omega_{k+1}={\textstyle \frac{1}{2}}\sum_{m\in {\mathbb Z}}\sum_{i=1}^d :h^i(-m)h_i(k+m): \; .\ee
Note that $\omega=L_{-2}1$.
The operator identity
\be [k(m),L_n]=mk(m+n)\label{lalala} \ee
implies that the $L_n$ form a Virasoro algebra of central charge $d$ 
(cf. \cite{KR}). Taking $n=-1$ we find   
\be Y(L_{-1}v,z)=\dz Y(v,z) \ee
so that $L_{-1}=D$.
Now let $v=h_1(-n_1)\cdots h_r(-n_r)\otimes e^{\alpha}$. Then 
\be L_0 v= (n_1+\dots+n_r+{\textstyle \frac{1}{2}}\alpha^2) v \, .\label{wsi}
\ee
Putting this together gives
\begin{pp1}
$\omega$ is a Virasoro element in V.
\end{pp1}
Equation (\ref{wsi}) shows that the spectrum of $L_0$ is in general unbounded
and that the eigenspaces of $L_0$ may be infinite-dimensional.

Let $y\in V$ and $x=k(-n)y$ with $n>0,\, k\in h$. Then iterating (\ref{lalala}) gives 
\be
L_1^qx = \sum_{p=0}^q {q\choose p}{n \choose q-p}(q-p)!k\big(-\{n-(q-p)\}\big)
                                                     L_1^py \,.
\label{aid} \ee
It follows
\begin{pp1}
$L_1$ is locally nilpotent on $V$.
\end{pp1}
{\it Proof:} Let $y=e^{\alpha}$. Then $L_1y=0$. We now prove by induction on $m$ that for $x=k_m(-n_m)\cdots k_1(-n_1)\otimes e^{\alpha}$ there is a $j$ such that $L^jx=0$. If $x=k_1(-n_1)\otimes e^{\alpha}$ then $L^{n_1}x=n_1!k_1(0)e^{\alpha}
=n_1!\la k_1,\alpha \ra e^{\alpha}$ by (\ref{aid}) and $L^{n_1+1}x=0$. Now let 
$x=k_m(-n_m)\cdots k_1(-n_1)\otimes e^{\alpha}$ and put 
$y=k_{m-1}(-n_{m-1})\cdots k_1(-n_1)\otimes e^{\alpha}$ such that $x=k_m(-n_m)y$. By induction hypothesis there is a $j$ such that $L^jy=0$. 
If we choose $q$ such that $q-j>n_m$ then (\ref{aid}) shows that $L_1^qx=0$.
\hspace*{\fill} $\Box$\\
 
We consider some examples. Directly from the definition we get
\be k(-1)_n=k(n) \ee
for $k\in h$ and $n\in {\mathbb Z}$. We will also need
\be
(1\otimes e^{\alpha})_n(1\otimes e^{\beta}) 
= \varepsilon(\alpha,\beta) S_{-n-1-\la \alpha,\beta \ra}
    (\alpha)\otimes e^{\alpha+\beta} \label{endlich} 
\ee
for $n\in {\mathbb Z},\, \alpha,\beta \in L$ . Here $S_{m}(\alpha)=S_{m}(\alpha(-1),\alpha(-2),\ldots,\alpha(-m))$ denotes the Schur polynomial which is defined by the generating function
\be 
\exp \left(\sum_{n\geq 1}\alpha (-n)\frac{z^n}{n} \right) = \sum_{m\geq 0} S_{m}(\alpha(-1),\alpha(-2),\ldots,\alpha(-m))z^m\, ,
\ee
for example,
\bea
 S_m(\alpha) &=& 0 \quad\mbox{for }\, m<0 \\
 S_0(\alpha) &=& 1 \\
 S_1(\alpha) &=& \alpha(-1).
\eea
Now the product is calculated as 
\beann
\lefteqn{ Y(1\otimes e^{\alpha},z)(1\otimes e^{\beta})} \\[3mm]
 &=& 
     \exp \left(\sum_{m\geq 1}\alpha (-m)\frac{z^m}{m} \right) 
     e^{\alpha}z^{\alpha(0)} 
     \exp \left(-\sum_{m\geq 1}\alpha (m)\frac{z^{-m}}{m} \right) 
     \varepsilon_{\alpha}  (1\otimes \beta) \\
 &=& \varepsilon(\alpha,\beta)
     \exp \left(\sum_{m\geq 1}\alpha (-m)\frac{z^m}{m} \right)e^{\alpha+\beta}
     z^{\la \alpha,\beta \ra} \\
 &=& \varepsilon(\alpha,\beta)
     \sum_{m\geq 0}S_{m}( \alpha) e^{\alpha+\beta}z^{m+\la \alpha,\beta \ra}
\eeann
The coefficient of $z^{-n-1}$ on the left hand side is $(1\otimes e^{\alpha})_n(1\otimes e^{\beta})$ and on the right hand side $\varepsilon(\alpha,\beta) S_{-n-1-\la \alpha,\beta \ra}(\alpha)\otimes e^{\alpha+\beta}$. This proves (\ref{endlich}). 
\beann
\lefteqn{ Y(h(-1)\otimes e^{\alpha},z)(1\otimes e^{\beta}) } \\
&=& : Y(h(-1)\otimes e^0,z)Y(1 \otimes e^{\alpha},z): 1\otimes e^{\beta} \\
&=& : \sum_{n\in \mathbb Z} h(n)z^{-n-1}
      \exp \big(\sum_{m\geq 1}\alpha (-m)\frac{z^m}{m} \big) 
      e^{\alpha}z^{\alpha(0)} \\
& &   \qquad\qquad
      \exp \big(-\sum_{m\geq 1}\alpha (m)\frac{z^{-m}}{m} \big) 
      \varepsilon_{\alpha}: 1\otimes e^{\beta} \\
&=&   : \sum_{n\in \mathbb Z} h(n)z^{-n-1}
      \exp \big(\sum_{m\geq 1}\alpha (-m)\frac{z^m}{m} \big) 
      e^{\alpha}z^{\alpha(0)}
      \varepsilon_{\alpha}: 1\otimes e^{\beta} \\
&=&  : h(0)z^{-1}\big( \sum_{m\geq 0}S_m({\alpha})z^m\big)
       e^{\alpha}z^{{\alpha}(0)}
        \varepsilon_{\alpha} : 1\otimes e^{\beta} \\
& & + :\sum_{n\geq 1}h(-n)z^{n-1}\big( \sum_{m\geq 0}S_m({\alpha})z^m\big)
       e^{\alpha}z^{{\alpha}(0)}
        \varepsilon_{\alpha}: 1\otimes e^{\beta} \\
&=&  \varepsilon(\alpha,\beta)\Big\{
     \sum_{m\geq 0}S_m({\alpha}) \la h,{\beta} \ra 
      z^{m+\la {\alpha},{\beta} \ra -1} \\
& &   \qquad\qquad
      +\sum_{n\geq 1}\sum_{m\geq 0}S_m({\alpha})h(-n)
      z^{m+n+ \la {\alpha},{\beta} \ra -1}\Big\}
       e^{{\alpha}+{\beta}}
\eeann
$(h(-n)\otimes e^{\alpha})_k(1\otimes e^{\beta})$ is the coefficient of $z^{-k-1}$ in this expression. Hence 
\be (h(-n)\otimes e^{\alpha})_k(1\otimes e^{\beta})=0 
\quad\mbox{for}\quad \la \alpha, \beta \ra -1\geq -k\, . \ee
In the other cases we get
\bea 
\lefteqn{ (h(-n)\otimes e^{\alpha})_k(1\otimes e^{\beta}) } \label{sgg} \\
&=& \varepsilon(\alpha,\beta)\Big\{ 
\la h,{\beta} \ra S_{-k-\la \alpha, \beta \ra}(\alpha)
+\sum_{m\geq 0}
S_m(\alpha)h(k+\la \alpha, \beta \ra +m)
\Big\} e^{{\alpha}+{\beta}} \nonumber
\eea

\section{Fermionic construction}
 
This construction is similar to the bosonic construction. Here we will work with the exterior algebra instead of the symmetric algebra since fermions are anticommuting.

Let $V$ be a real vector space. Define $T^p(V)=V\otimes \cdots \otimes V$ and the tensor algebra $T(V)=\bigoplus_{p\geq 0}T^p(V)$ of $V$. If $I$ is the ideal of $T(M)$ generated by all elements $v\otimes v$ with $v\in V$, then the exterior algebra $\wedge(V)$ of $V$ is defined by
\be  \wedge (V) =T(V)/ I. \ee
Then 
\be \wedge (V) =\bigoplus_{p\geq 0} \wedge^p (V) \ee
with $\wedge^p (V)=T^p(V)/(I\cap T^p(V))$. The product in $\wedge(V)$ is graded commutative, i.e.
\be 
vw=(-1)^{pq}wv \quad \mbox{for } v\in \wedge^p (V),\, w\in \wedge^q (V). 
\ee

Let $A$ be a finite-dimensional real vector space of dimension $d$ with a nondegenerate symmetric bilinear form $\la\, ,\, \ra$. We define the infinite-dimensional Heisenberg superalgebra
\be 
\hat{A}=A\otimes {\mathbb R}[t,t^{-1}]t^{\frac{1}{2}}\oplus {\mathbb R}c 
\ee
with even part 
\be \hat{A}_{\ov{0}}={\mathbb R}c \ee
and odd part
\be \hat{A}_{\ov{1}}=A\otimes {\mathbb R}[t,t^{-1}]t^{\frac{1}{2}} \ee
and products
\bea 
 & [a(m),b(n)]_+=\la a,b \ra \delta_{m+n,0}c & \\
 & [a(m),c]=[c,c]=0\, .                          &
\eea 
Then 
\be \hat{A}^-=A\otimes {\mathbb R}[t^{-1}]t^{-\frac{1}{2}} \ee
is an anticommuting subalgebra of $\hat{A}$, and $V=\wedge (\hat{A}^-)$ the exterior algebra of $\hat{A}^-$. Let 
\bea
 V_{\ov{0}} &=& \bigoplus_{n\in 2{\mathbb Z} \atop n\geq 0} \wedge^n (\hat{A}^-) \\
 V_{\ov{1}} &=& \bigoplus_{n\in 2{\mathbb Z}+1 \atop n\geq 0} \wedge^n (\hat{A}^-) \, .
\eea
Then 
\be V=V_{\ov{0}}\oplus V_{\ov{1}}\, . \ee
$\hat{A}$ acts canonically on $V$ by 
\bea
 a(n)\cdot v &=& \partial_{a(-n)}v \\
 a(-n)\cdot v &=& a(-n)v \\
    c\cdot  v &=& v
\eea
for all $a\in \hat{A},\, v\in V$ and $n>0,\, n\in {\mathbb Z}+\frac{1}{2}$. Here the operator $\partial_{a(-n)}$ is defined by
\bea
\partial_{a(-n)} 1     &=& 0 \\
\partial_{a(-n)} b(-m) &=& \la a,b \ra \delta_{mn} \\
\partial_{a(-n)} (vw)  &=& (\partial_{a(-n)}v)w+(-1)^{|v|}v(\partial_{a(-n)}w)
\, .
\eea
For $1\in {\mathbb R}=\wedge^0 (\hat{A}^-)\subset V_{\ov{0}}$ put 
\be Y(1,z)=1 \, . \ee
For $a\in A,\, 0\leq n\in {\mathbb Z}$ let
\be Y(a(-n-{\textstyle \frac{1}{2}}),z)=\frac{1}{n!}\left( \dz \right)^n \left(\frac{a(z)}{z^{\frac{1}{2}}} \right) \ee
where
\be a(z)=\sum_{m\in {\mathbb Z}+\frac{1}{2}}a(m)z^{-m}\, . \ee
We define the fermionic normal ordering
\be 
{\textstyle {\circ \atop \circ}} 
a_1(n_1)\cdots a_r(n_r)
{\textstyle {\circ \atop \circ}} \: = 
\mbox{sgn}(\sigma) a_{\sigma 1}(n_{\sigma 1})\cdots a_{\sigma r}(n_{\sigma r})
\ee
 for $a_1,\ldots ,a_r\in A,\, n_1,\ldots ,n_r\in {\mathbb Z}+\frac{1}{2}$, where $\sigma \in S_r$ is any permutation such that $n_{\sigma 1}\leq \ldots \leq n_{\sigma r}$. Note that $
{\textstyle {\circ \atop \circ}} 
a_1(n_1)a_2(n_2)
{\textstyle {\circ \atop \circ}} 
\: = \:
-{\textstyle {\circ \atop \circ}} 
a_2(n_2)a_1(n_1)
{\textstyle {\circ \atop \circ}} \: .
$

For $0\leq n_1,\ldots ,n_r\in {\mathbb Z}$ and 
\be v=a_1(-n_1-{\textstyle \frac{1}{2}})\cdots a_r(-n_r-{\textstyle \frac{1}{2}}) \ee
define 
\be
Y(v,z)=
{\textstyle {\circ \atop \circ}} 
\: Y(a(-n_1-{\textstyle \frac{1}{2}}),z)
     \cdots Y(a(-n_r-{\textstyle \frac{1}{2}}),z) \:
{\textstyle {\circ \atop \circ}}
\, .
\ee
As in the bosonic case this gives us a well-defined linear map
\beann 
Y\; :\; V &\rightarrow & End(V)[\![z,z^{-1}]\!] \\
        v &\mapsto     & Y(v,z)=\sum_{n\in {\mathbb Z}}v_nz^{-n-1}, 
                                 \quad v_n\in End(V)
\eeann
and
\begin{pp1} 
Y provides $V=V_{\ov{0}}\oplus V_{\ov{1}}$ with the structure of a vertex superalgebra. V is simple. 
\end{pp1} 

It is easy to see that $V$ contains a Virasoro element. Let $\{b^1,\ldots,b^d\}$ be a basis of $A$ and $\{b_1,\ldots,b_d\}$ the corresponding dual basis. The element 
\be \omega={\textstyle \frac{1}{2}} \sum_{i=1}^d 
b^i(-{\textstyle \frac{3}{2}})b_i(-{\textstyle \frac{1}{2}}) \ee
is independent of the choice of basis of $A$. Writing 
\[ \omega={\textstyle \frac{1}{4}} \sum_{i=1}^d \Big(
  b^i(-{\textstyle \frac{3}{2}}) b_i(-{\textstyle \frac{1}{2}}) 
+ b_i(-{\textstyle \frac{3}{2}}) b^i(-{\textstyle \frac{1}{2}}) \Big) \]
gives
\beann
 \lefteqn{Y(\omega,z)} \\
&=& {\textstyle \frac{1}{4}}\sum_{i=1}^d \Big( 
Y(b^i(-{\textstyle \frac{3}{2}}) b_i(-{\textstyle \frac{1}{2}}),z)+
Y(b_i(-{\textstyle \frac{3}{2}}) b^i(-{\textstyle \frac{1}{2}}),z) \Big) \\
&=& -{\textstyle \frac{1}{4}}\sum_{i=1}^d \Big(
 \sum_{m,n \in \mathbb Z+\frac{1}{2}} 
 (m+{\textstyle \frac{1}{2}})\fn b^i(m)b_i(n) \fn \: z^{-(m+n)-2} \\
& &
+ \sum_{m,n \in \mathbb Z+\frac{1}{2}}
 (n+{\textstyle \frac{1}{2}})\fn b_i(n)b^i(m) \fn  \: z^{-(m+n)-2} \Big) \\
&=& -{\textstyle \frac{1}{4}}\sum_{i=1}^d 
 \sum_{m,n \in \mathbb Z+\frac{1}{2}}
  (m-n)\fn b^i(m)b_i(n) \fn \: z^{-(m+n)-2} \\
&=& {\textstyle \frac{1}{2}}\sum_{i=1}^d 
 \sum_{k \in \mathbb Z} \sum_{n \in \mathbb Z+\frac{1}{2}}
 (n-{\textstyle \frac{k}{2}}) \fn b^i(k-n)b_i(n) \fn \: z^{-k-2} \\
&=& {\textstyle \frac{1}{2}}\sum_{i=1}^d 
 \sum_{k \in \mathbb Z} \sum_{n \in \mathbb Z+\frac{1}{2}}
 (n+{\textstyle \frac{k}{2}}) \fn b_i(-n)b^i(k+n) \fn \: z^{-k-2} \\
&=& \sum_{k \in \mathbb Z} L_k z^{-k-2}
\eeann
with
\be L_k= \omega_{k+1}= {\textstyle \frac{1}{2}} 
        \sum_{n \in {\mathbb Z}+\frac{1}{2}}
        \sum_{i=1}^d (n+{\textstyle \frac{k}{2}}) \fn b_i(-n)b^i(k+n) \fn \, .
\ee
From the operator identity
\be [a(m),L_n]=(m+{\textstyle \frac{n}{2}})a(m+n) \label{joho} \ee
follows that the $L_n$ form a Virasoro algebra of central charge 
$c=\frac{d}{2}$ (cf. \cite{KR} p. 29f).
Let $v=a_1(-m_1)\ldots a_r(-m_r),\, m_i>0, \,a_i\in A$. Then (\ref{joho}) yields (cf. \cite{FFR})
\be Y(L_{-1}v,z)=\dz Y(v,z) \ee
and 
\be L_0 v= (m_1+\ldots+ m_r)v \label{lof2} \ee
so that 
\begin{pp1} 
$\omega$ is a Virasoro element in V.
\end{pp1}
Formula (\ref{lof2}) has some important consequences. It shows that the $V_n$ are finite-dimensional, $V_0={\mathbb R}1$ and $V_n=0$ for $n<0$. This implies that $L_1$ is locally nilpotent on $V$.

We will need the following formulas
\bea
a({\textstyle -\frac{1}{2}})_n &=& a(n+{\textstyle \frac{1}{2}}) \\
\big( a({\textstyle -\frac{1}{2}})b({\textstyle -\frac{1}{2}})\big)_n 
&=& \sum_{2m\leq n-1}a(m+{\textstyle \frac{1}{2}})
                     b(n-m-{\textstyle \frac{1}{2}}) \nonumber \\
& & -\sum_{2m>n-1}b(n-m-{\textstyle \frac{1}{2}})
                  a(m+{\textstyle \frac{1}{2}}) \label{schellistdoof}  \, .  
\eea
The first one follows directly from the definition.
\beann
Y(a({\textstyle -\frac{1}{2}})b({\textstyle -\frac{1}{2}}),z)
&=& \fn 
        \sum_{m\in \mathbb Z}a(m+{\textstyle \frac{1}{2}})z^{-m-1}
        \sum_{k\in \mathbb Z}b(k+{\textstyle \frac{1}{2}})z^{-k-1} \fn \\
&=& \sum_{k,m\in \mathbb Z}
      \fn a(m+{\textstyle \frac{1}{2}})
          b(k+{\textstyle \frac{1}{2}}) \fn \: z^{-(k+m+1)-1} \\
&=& \sum_{n\in \mathbb Z}\sum_{m\in \mathbb Z} 
       \fn a(m+{\textstyle \frac{1}{2}})
           b(n-m-{\textstyle \frac{1}{2}}) \fn \: z^{-n-1} 
\eeann
so that 
\beann
\lefteqn{ 
\big( a({\textstyle -\frac{1}{2}})b({\textstyle -\frac{1}{2}})\big)_n } \\
&=& 
\sum_{2m\leq n-1}a(m+{\textstyle \frac{1}{2}})
                     b(n-m-{\textstyle \frac{1}{2}})
-\sum_{2m>n-1}b(n-m-{\textstyle \frac{1}{2}})
                  a(m+{\textstyle \frac{1}{2}}) \, .
\eeann

\chapter[The Neveu-Schwarz superstring]{The vertex superalgebra of the compactified Neveu-Schwarz superstring and Lie algebras}

In this chapter we introduce the vertex superalgebra of the compactified Neveu-Schwarz superstring and show that it contains a Neveu-Schwarz element. Thus the quotient $G=G_{-\frac{1}{2}}\mbox{\sf P}_{\frac{1}{2}}/D\mbox{\sf P}_0$ of the physical states $G_{-\frac{1}{2}}\mbox{\sf P}_{\frac{1}{2}}$ and the null states $D\mbox{\sf P}_0$ is a Lie algebra. We study this Lie algebra in detail. It contains sometimes a Kac-Moody algebra generated by the tachyon and the first excited states. The states invariant under the GSO-projection still form a Lie algebra $G^+$. Then we construct an invariant form on $G$ and $G^+$ and prove that the quotients of $G$ and $G^+$ with the kernel of this form are generalized Kac-Moody algebras in the case of Lorentzian space-time. Furthermore we determine the roots and calculate their multiplicities for space-time dimensions smaller than or equal to $10$. 

For physical background consider \cite{GSW}.

\section{Definition and some properties} \label{cpnss}

Let $L$ be an integral nondegenerate lattice of rank $d$ and $\alpha^{\mu},\,
1\leq \mu \leq d$ a basis of $h=L\otimes_{\mathbb Z}{\mathbb R}$. Define 
the matrix $\eta^{\mu \nu}=\la \alpha^{\mu},\alpha^{\nu} \ra $ and the inverse 
$\eta_{\kappa \lambda}$, i.e. $\eta^{\mu \nu} \eta_{\nu \lambda}=\delta^{\mu}_{\lambda}$.
Denote the bosonic vertex superalgebra corresponding to $L$ by $V_{bos}$.

Let $A$ be an isomorphic copy of $h$ provided with the same nondegenerate 
bilinear form $\la \, ,\, \ra$ and with basis $b^{\mu},\, 1\leq \mu \leq d$.
The fermionic vertex superalgebra constructed from $A$ is called $V_{ferm}$.

The vertex superalgebra of the compactified Neveu-Schwarz superstring is 
defined to be 
\be V=V_{ferm} \otimes V_{bos}. \ee 
$V$ is simple since $V_{ferm}$ and $V_{bos}$ are.
The even part of $V$ is 
\be V_{\ov{0}}= 
\bigoplus_{n\in 2{\mathbb Z}\atop n\geq 0} \wedge^n (\hat{A}^-)
\otimes S(\hat{h}^-)\otimes {\mathbb R}[L_0]
\oplus 
\bigoplus_{n\in 2{\mathbb Z}+1 \atop n\geq 0} \wedge^n (\hat{A}^-)
\otimes S(\hat{h}^-)\otimes {\mathbb R}[L_1] 
\ee 
and the odd part 
\be V_{\ov{1}}=
\bigoplus_{n\in 2{\mathbb Z}+1 \atop n\geq 0} \wedge^n (\hat{A}^-)
\otimes S(\hat{h}^-)\otimes {\mathbb R}[L_0]
\oplus 
\bigoplus_{n\in 2{\mathbb Z} \atop n\geq 0} \wedge^n (\hat{A}^-)
\otimes S(\hat{h}^-)\otimes {\mathbb R}[L_1] \, .
\ee
$\omega=1\otimes \omega_{bos}+\omega_{ferm}\otimes 1$ is a Virasoro element in $V$. The corresponding Virasoro 
ope\-rators are given by 
\be L_m = L_m^{bos}+L_m^{ferm} \ee
with
\bea
L_m^{bos} &=& {\textstyle\frac{1}{2}} \sum_{n\in \mathbb Z} 
               :\alpha (-n)\cdot \alpha (m+n) : \\
L_m^{ferm} &=& {\textstyle\frac{1}{2}} \sum_{r\in \mathbb Z+ \frac{1}{2}}
              (r+{\textstyle\frac{1}{2}}m)\fn b(-r)\cdot b(m+r)\fn 
\eea
where we have written $\alpha (-n)\cdot \alpha (m+n)$ for 
$\eta_{\mu \nu}\alpha^{\mu} (-n) \alpha^{\nu} (m+n)$ and similar for $b$. 

Let $v=a_1(r_1)\ldots a_p(r_p)\otimes h_1(n_1)\ldots h_q(n_q)\otimes e^{\alpha}$ with $a_i\in A,\, h_j\in h$ and $\alpha \in L$ be in $V$. Then 
\be 
L_0 v = (r_1+\ldots +r_p+n_1+\ldots +n_q+{\textstyle\frac{1}{2}}\alpha^2 )v\, .
\ee

Furthermore
\begin{p1}
$\tau= b(-\frac{1}{2}) \cdot \alpha(-1) \in V_{\ov{1}}$ is a Neveu-Schwarz 
element of central charge $\frac{3}{2}d$.
\end{p1}
{\it Proof:}
The vertex operator of $\tau$ is
\beann
Y(\tau,z) &=& Y(b(-{\textstyle{\frac{1}{2}}}),z)\cdot Y(\alpha(-1),z) \\
          &=& \eta_{ \mu \nu }
              \Big( \sum_{r\in \mathbb Z +\frac{1}{2}} b^{\mu}(r)
                          z^{-r-\frac{1}{2}}\Big) 
              \Big( \sum_{n\in \mathbb Z } \alpha^{\nu}(n)
                          z^{-n-1 }\Big) \\
 &=& \eta_{ \mu \nu }\sum_{r\in \mathbb Z +\frac{1}{2}}\sum_{n\in \mathbb Z }
     \alpha^{\nu}(n)b^{\mu}(r)z^{-(r+n)-\frac{3}{2} } \\
 &=& \sum_{r\in \mathbb Z +\frac{1}{2}}G_{r}z^{-r-\frac{3}{2} } 
\eeann
with 
\be G_{r}=\tau_{r+\frac{1}{2}}= \sum_{n\in \mathbb Z} 
          \alpha(-n)\cdot b(r+n) \ee

Note that $\tau=G_{-\frac{3}{2}}1$.

Furthermore
\beann 
{\textstyle\frac{1}{2}}{\tau}_0\tau 
  &=& {\textstyle\frac{1}{2}}
               G_{-\frac{1}{2}}\tau \\
  &=& {\textstyle\frac{1}{2}} \Big( \sum_{n\in \mathbb Z}
         \alpha (-n)\cdot b(n-{\textstyle\frac{1}{2}}) \Big)\;
                     b(-{\textstyle\frac{1}{2}})\cdot \alpha(-1) \\
  &=& {\textstyle\frac{1}{2}} 1\otimes \alpha(-1)\cdot 
                     \alpha(-1) +
      {\textstyle\frac{1}{2}} b(-{\textstyle\frac{3}{2}})\cdot
                     b(-{\textstyle\frac{1}{2}})\otimes 1 \\
  &=& 1\otimes \omega_{bos}+\omega_{ferm}\otimes 1  \, .
\eeann
It is well known that the operators $L_n,\, n\in {\mathbb Z}$ and $G_r,\, r\in {\mathbb Z}+\frac{1}{2}$ give a representation of the Neveu-Schwarz algebra with central charge $\frac{3}{2}d$ (cf. \cite{GSW} vol.1 p.204). \hspace*{\fill} $\Box$\\

We decompose $V$ in $V=V^+\oplus V^-$ where 
\be 
V^+=\bigoplus_{n\in 2{\mathbb Z}+1 \atop n\geq 0} \wedge^n (\hat{A}^-)
\otimes S(\hat{h}^-)\otimes {\mathbb R}[L_1]\oplus \bigoplus_{n\in 2{\mathbb Z}+1 \atop n\geq 0} \wedge^n (\hat{A}^-)
\otimes S(\hat{h}^-)\otimes {\mathbb R}[L_0]
\ee
is the subspace of $V$ with an odd number of fermionic oscillator excitations
and 
\be V^-=\bigoplus_{n\in 2{\mathbb Z}\atop n\geq 0} \wedge^n (\hat{A}^-)
\otimes S(\hat{h}^-)\otimes {\mathbb R}[L_0]
\oplus 
\bigoplus_{n\in 2{\mathbb Z} \atop n\geq 0} \wedge^n (\hat{A}^-)
\otimes S(\hat{h}^-)\otimes {\mathbb R}[L_1]
\ee
the subspace with an even number of fermionic oscillator excitations.
Define the GSO-projection $G$ on $V$ by putting $G=+1$ on $V^+$ and $G=-1$
on $V^-$. It is easy to see that $G$ commutes with the operators $k(n),\, k\in h,\, n\in {\mathbb Z},$ and anticommutes with the operators $a(r),\, a\in A,\, r\in {\mathbb Z}+\frac{1}{2},$ so that  
\be [G,L_n]={[}G,G_r{]_+}=0 \ee
for all $n\in {\mathbb Z}$ and all $r\in {\mathbb Z}+\frac{1}{2}$. This implies
\be 
L_nV^{\pm} \subset V^{\pm} \quad \mbox{and} \quad 
G_rV^{\pm} \subset V^{\mp}\, .
\ee
Finally 
\bea
(V^+)_n(V^+) &\subset & V^- \\
(V^+)_n(V^-) &\subset & V^+ \\
(V^-)_n(V^-) &\subset & V^- \\
(V^-)_n(V^+) &\subset & V^+ \, .
\eea

\section{Physical states and Lie algebras} \label{bigwaves}

$\mbox{\sf P}_{\frac{1}{2}}=\{v\in V| L_0v={\textstyle\frac{1}{2}}v,\, 
L_mv=G_rv=0\, \mbox{ for all } m,r> 0 \}$ is called the space of physical states 
of the compactified Neveu-Schwarz superstring. We will give some examples for 
physical states and calculate their vertex operators. The state $|k\ra =1\otimes 1\otimes e^k$ with $k^2=1$ is a ground-state tachyon. 
The first excited state is the massless vector $\zeta(-\frac{1}{2}) |k\ra 
=\zeta (-\frac{1}{2}) \otimes 1 \otimes e^k$ of polarisation $\zeta\in A$ and momentum $k^2=0$. The condition $G_{\frac{1}{2}}\, \zeta (-\frac{1}{2}) |k\ra =0$ implies that $\la \zeta , k \ra=0$ for this state to be physical. We have already seen that the physical states form the Lie algebra $G_{-\frac{1}{2}}\mbox{\sf P}_{\frac{1}{2}}/D\mbox{\sf P}_0$.
The vertex operator of $G_{-\frac{1}{2}}|k\ra =k(-\frac{1}{2})|k\ra $ is given by 
\be 
Y(G_{-\frac{1}{2}}|k\ra ,z)=\Big( \sum_{r\in \mathbb Z +\frac{1}{2}}
k(r)z^{-r-\frac{1}{2}} \Big) :e^{ikX(z)}: \varepsilon_k \, (-1)^{|.|} \ee
which is in agreement with the expression in \cite{GSW}. Note the sign operator acting on the fermionic part. For the first excited state we obtain $G_{-\frac{1}{2}}\zeta(-\frac{1}{2}) |k\ra = 
\left( \zeta(-1)-\zeta(-\frac{1}{2})k(-\frac{1}{2})\right)|k\ra $ and 
\pagebreak
\bea
\lefteqn{ 
Y\big(G_{{\textstyle -\frac{1}{2}}}\zeta({\textstyle-\frac{1}{2}}) |k\ra 
,z\big) } \\
&=& : \Big(\sum_{m\in \mathbb Z}\zeta(m)z^{-m-1}\Big) e^{ikX(z)}: \varepsilon_k \nonumber \\
&& - \,
\fn {\displaystyle 
    \Big(\sum_{r\in {\mathbb Z}+\frac{1}{2}}\zeta(r)z^{-r-\frac{1}{2}}\Big) 
    \Big(\sum_{s\in {\mathbb Z}+\frac{1}{2}}k(s)z^{-s-\frac{1}{2}}\Big)
    } 
\fn
\, :e^{ikX(z)}: \varepsilon_k \nonumber
\eea
Again this coincides with the result in \cite{GSW}.

We now can calculate some products in the Lie algebra $G=G_{-\frac{1}{2}}\mbox{\sf P}_{\frac{1}{2}}/D\mbox{\sf P}_0$. Let $|k\ra $ be a ground-state tachyon.
Then 
\beann
\big( G_{-\frac{1}{2}}|k\ra \big)_0\big( |-k\ra \big) 
&=&
\big( k({\textstyle -\frac{1}{2}})\otimes (1\otimes e^k)\big)_0
\big( 1 \otimes ( 1\otimes e^{-k}) \big) \\
&=&
\sum_{n\in \mathbb Z} k({\textstyle -\frac{1}{2}})_n1 \otimes 
(1\otimes e^k)_{-n-1}(1\otimes e^{-k}) \\
&=&
\sum_{n\in \mathbb Z} k(n+{\textstyle \frac{1}{2}})\, 1 \otimes
\varepsilon(k,-k)S_{n+1}(k)e^0 \\
&=& 
\varepsilon(k,-k) k({\textstyle -\frac{1}{2}})|0\ra
\eeann
and 
\be
[ G_{-\frac{1}{2}}|k\ra,G_{-\frac{1}{2}}|-k\ra ]
= -G_{-\frac{1}{2}} k({\textstyle -\frac{1}{2}})|0\ra \, . \ee
Let $\zeta(-\frac{1}{2})|0 \ra$ be an excited state of zero momentum. Then
\beann
\big( G_{-\frac{1}{2}}\zeta({\textstyle -\frac{1}{2}}) |0 \ra \big)_0
\big(|k\ra \big) 
&=&
\big( 1\otimes (\zeta(-1)\otimes 1) \big)_0 \big( 1\otimes (1\otimes e^k)\big) \\
&=&
\sum_{n\in \mathbb Z}1_n1 \otimes (\zeta(-1)\otimes 1)_{-n-1}(1\otimes e^k) \\
&=&
1\otimes (\zeta(-1)\otimes 1)_0(1\otimes e^k) \\
&=&
1 \otimes \zeta(0) (1\otimes e^k) \\
&=&
\la \zeta, k\ra |k\ra 
\eeann
and 
\be 
[ G_{-\frac{1}{2}}\zeta({\textstyle -\frac{1}{2}})|0\ra,G_{-\frac{1}{2}} |k\ra ]
= \la \zeta, k\ra G_{-\frac{1}{2}} |k\ra \, .
\ee
Furthermore
\beann
\big( G_{-\frac{1}{2}}\zeta({\textstyle -\frac{1}{2}})|0\ra \big)_0
\big( \xi({\textstyle -\frac{1}{2}})|0\ra \big) 
&=& 
\big( 1\otimes (\zeta(-1)\otimes 1) \big)_0 
\big( \xi({\textstyle -\frac{1}{2}}) \otimes (1\otimes e^0 ) \big) \\
&=& 
\sum_{n\in \mathbb Z} 1_n \xi({\textstyle -\frac{1}{2}}) \otimes
(\zeta(-1)\otimes 1)_{-n-1} 1 \\
&=&
0
\eeann
so that 
\be [G_{-\frac{1}{2}}\zeta({\textstyle -\frac{1}{2}})|0\ra ,
     G_{-\frac{1}{2}}\xi({\textstyle -\frac{1}{2}})|0\ra ]=0\, . 
\label{isww} \ee
$V$ is $L$-graded by construction
\be V=\bigoplus_{\alpha \in L} V^{\alpha} \ee
where
\be V^{\alpha} = \bigoplus_{n\geq 0}\wedge^n (\hat{A}^-)
\otimes S(\hat{h}^-)\otimes e^{\alpha} \, . \ee
Using the fact that the Neveu-Schwarz operators leave the spaces 
$V^{\alpha}$ invariant it is easy to see that
the subspaces $V_n$ and $\mbox{\sf P}_n$ are also $L$-graded, i.e. $V_n=\oplus_{\alpha \in L} V_n^{\alpha}$ where $V_n^{\alpha}=V_n \cap V^{\alpha}$ and analogous for $\mbox{\sf P}_n$. If $T$ is a Neveu-Schwarz operator then the spaces $TV_n$ are $L$-graded and $TV_n\cap V^{\alpha}=(TV_n)^{\alpha}=TV_n^{\alpha}=T(V_n\cap V^{\alpha})$. The same holds for the $\mbox{\sf P}_n$.  

It follows
\be G=\bigoplus_{\alpha \in L} G^{\alpha} \ee
with
\be  G^{\alpha}=(G_{-\frac{1}{2}}\mbox{\sf P}_{\frac{1}{2}})^{\alpha }/
                (D\mbox{\sf P}_0)^{\alpha}=G \cap V^{\alpha }\, . \ee
From the definition of the vertex operators we see
\be \big( V^{\alpha} \big)_n\big( V^{\beta} \big)\subset V^{\alpha + \beta} \ee
so that
\be [G^{\alpha},G^{\beta}] \subset G^{\alpha + \beta}\, . \ee
We now can embed a Kac-Moody algebra into $G$. Let $\Pi = \{ k_i\, |\, i\in I\}$ be a subset of $L$ satisfying
\begin{enumerate}
\item $\la k_i,k_i \ra = 1$ for all $i\in I$
\item $\la k_i,k_j \ra \leq 0 $ for $i\neq j$
\item $\Pi$ is linearly independent in $L_{\mathbb R}=
                                 L\otimes_{\mathbb Z}{\mathbb R}$
\end{enumerate}
Note that $\Pi$ is necessarily finite. We can choose nontrivial $\Pi$ for example if $L$ is the odd Lorentzian lattice $I_{n,1}$. For $i,j\in I$ define
\bea
a_{ij} &=& 2 \la k_i,k_j \ra \\
h_i    &=& 2G_{-\frac{1}{2}}k_i({\textstyle -\frac{1}{2}})|0\ra \\
e_i    &=& \sqrt{2}G_{-\frac{1}{2}}|k_i\ra \\
f_i    &=& -\sqrt{2}G_{-\frac{1}{2}}|-k_i\ra \, .
\eea
\begin{pp1}
The elements $e_i,f_i$ and $h_i$ generate the Kac-Moody algebra $g(A)$ corresponding to the matrix $A=(a_{ij})$.
\end{pp1}
{\it Proof:} We have already established the relations
\beann
{[}h_i,h_j{]} &=& 0 \\
{[}h_i,e_j{]} &=& a_{ij}e_j \\
{[}h_i,f_j{]} &=& -a_{ij}f_j \\
{[}e_i,f_i{]} &=& h_i \, .
\eeann
Let $i\neq j$. Then
\beann
{[}e_i,f_j{]} &=& -2 [ G_{-\frac{1}{2}}|k_i\ra,G_{-\frac{1}{2}}|-k_j\ra]\\
&=& 2\big( k_i({\textstyle -\frac{1}{2}})\otimes (1\otimes e^{k_i})\big)_0
     \big( k_j({\textstyle -\frac{1}{2}})\otimes (1\otimes e^{-k_j})\big) \\
&=& -2\sum_{n\in \mathbb Z}
 k_i({\textstyle -\frac{1}{2}})_nk_j({\textstyle -\frac{1}{2}}) \otimes
 (1\otimes e^{k_i})_{-n-1}(1\otimes e^{-k_j}) \\
&=& -2\sum_{n\in \mathbb Z}
k_i({\textstyle n+\frac{1}{2}})k_j({\textstyle -\frac{1}{2}}) \otimes
\varepsilon(k_i,-k_j)S_{n+\la k_i,k_j \ra}(k_i)e^{k_i-k_j}\\
&=& 
-2 \varepsilon(k_i,-k_j) \la k_i,k_j \ra S_{\la k_i,k_j \ra}(k_i)e^{k_i-k_j}\\
&=& 0
\eeann
since $\la k_i,k_j \ra \leq 0$. Finally we must prove the Serre relations
\[ 
(ad\, e_i)^{1-a_{ij}}e_j=(ad\, f_i)^{1-a_{ij}}f_j=0\quad\mbox{for }i\neq j. \]
$(ad\, e_i)^ne_j$ can be written as $(ad\, e_i)^ne_j=G_{-\frac{1}{2}}x$ modulo $D\mbox{\sf P}_0$ where $x$ is some element in $\mbox{\sf P}_{\frac{1}{2}}$ which is of degree $nk_i+k_j$ with respect to the $L$-grading. Thus the $L_0$-eigenvalue of $x$ is greater or equal to $\frac{1}{2}(nk_i+k_j)^2=\frac{1}{2}(n^2+2n \la k_i,k_j \ra + 1)$.
If $\frac{1}{2}(n^2+na_{ij} + 1)>\frac{1}{2}$ then $x$ must be zero. We conclude that $(ad\, e_i)^ne_j=0$ for $n>-a_{ij}$ resp. $n\geq 1-a_{ij}$. The proof of the other Serre relation is analogous. 
\hspace*{\fill} $\Box$\\

Since $g(A)$ is a subalgebra of $G$ it is also $L$-graded, i.e.
\be g(A)=\bigoplus_{\alpha \in L} g(A)^{\alpha} \, .\ee
An element $\alpha\in L\backslash \{0\}$ with $g(A)^{\alpha}\neq 0$ is         called a root of $g(A)$. The inclusion $g(A)^{\alpha}\subset G^{\alpha}$
implies that the dimension of a root space $g(A)^{\alpha}$ is bounded by the dimension of $G^{\alpha}$ which can be calculated.
Define numbers $c_d(n)$ by
\be 
\sum_{n=0}^{\infty} c_d(n)q^n = \prod_{m=1}^{\infty}
     \left( \frac{1+q^{m-\frac{1}{2}}}{1-q^m} \right)^d  \, .   \ee
The first nonzero coefficients are
\beann 
c_d(0)                        &=& 1 \\
c_d({\textstyle \frac{1}{2}}) &=& d \\
c_d(1)                        &=& d+{\textstyle {d \choose 2}} \\
c_d({\textstyle \frac{3}{2}}) &=& d+d^2+{\textstyle {d \choose 3}} \\
c_d(2)                        &=& 2d+d^2+d{\textstyle {d \choose 2}}
                      +{\textstyle {d \choose 2}}+{\textstyle {d \choose 4}}
                                                                  \, .
\eeann
We have seen above that $G_{-\frac{1}{2}}$ is injective on $\mbox{\sf P}_{\frac{1}{2}}$ and that $D$ is injective on $\mbox{\sf P}_0^{\alpha}$. Hence 
$\mbox{dim }{G}^{\alpha}=\mbox{dim }\mbox{\sf P}_{\frac{1}{2}}^{\alpha}-
\mbox{dim }\mbox{\sf P}_0^{\alpha}$. 
We prove in the appendix 
\be 
\mbox{dim} \, \mbox{\sf P}^{\alpha}_{n}=
               c_{d-1}(n-{\textstyle \frac{1}{2}}{\alpha}^2)\, , \label{ahd}
\ee
where $d$ denotes the rank of the lattice,
so that we get the following upper bound for the multiplicities
\be 
\mbox{dim }g(A)^{\alpha}\leq c_{d-1}({\textstyle \frac{1}{2}}(1-{\alpha}^2))
                            -c_{d-1}(-{\textstyle \frac{1}{2}}{\alpha}^2)\, .
\ee

With the help of the GSO-projection we construct another subalgebra of $G$.
Put \be
\mbox{\sf P}_{n}^{\pm}=\mbox{\sf P}_{n}\cap V^{\pm} \quad \mbox{for }n\in 
{\textstyle \frac{1}{2}}{\mathbb Z}\, . \ee 
Let $x=x^++x^-\, , x^{\pm}\in V^{\pm}$ be in $\mbox{\sf P}_{n}$. Then $L_0x^++L_0x^-=nx^++nx^-$ and by applying $G$ on this equation $L_0x^+-L_0x^-=nx^+-nx^-$ so that $L_0x^{\pm}=nx^{\pm}$.      
For $r>0$ we have $G_{r}x^++G_{r}x^-=0$ and 
$-G_{r}x^++G_{r}x^-=0$ by applying $G$. Thus $G_rx^{\pm}=0$. This shows 
$x^{\pm}\in \mbox{\sf P}_{n}^{\pm}$ and 
\be  \mbox{\sf P}_{n}=\mbox{\sf P}_{n}^+\oplus \mbox{\sf P}_{n}^- \, . \ee
\begin{pp1}
$G_{-\frac{1}{2}}\mbox{\sf P}_{\frac{1}{2}}^+/D\mbox{\sf P}_0^-$ is a subalgebra of $G_{-\frac{1}{2}}\mbox{\sf P}_{\frac{1}{2}}/D\mbox{\sf P}_0$.
\end{pp1}
{\it Proof:} Let $u,v\in \mbox{\sf P}_{\frac{1}{2}}^+$. Then $(G_{-\frac{1}{2}}u)_0v$ is in $\mbox{\sf P}_{\frac{1}{2}}\cap V^+=\mbox{\sf P}_{\frac{1}{2}}^+$ and $(G_{-\frac{1}{2}}u)_0(G_{-\frac{1}{2}}v)\linebreak =
     G_{-\frac{1}{2}}((G_{-\frac{1}{2}}u)_0v)$ implies that 
      $\big( G_{-\frac{1}{2}}\mbox{\sf P}_{\frac{1}{2}}^+ \big)_0
       \big( G_{-\frac{1}{2}}\mbox{\sf P}_{\frac{1}{2}}^+ \big)\subset
       \big( G_{-\frac{1}{2}}\mbox{\sf P}_{\frac{1}{2}}^+ \big)$. 
It remains to show that $G_{-\frac{1}{2}}\mbox{\sf P}_{\frac{1}{2}}^+
\cap D\mbox{\sf P}_0 = D\mbox{\sf P}_0^-$. $G_{-\frac{1}{2}}\mbox{\sf P}_0^-$ is a subspace of 
$\mbox{\sf P}_{\frac{1}{2}}$ since $G_{-\frac{1}{2}}\mbox{\sf P}_0$ is. But 
$G_{-\frac{1}{2}}\mbox{\sf P}_0^-$ is also in $V^+$. Thus $G_{-\frac{1}{2}}\mbox{\sf P}_0^-\subset\mbox{\sf P}_{\frac{1}{2}}^+$ and $G_{-\frac{1}{2}}\mbox{\sf P}_0^-\subset ( G_{-\frac{1}{2}}\mbox{\sf P}_0 \cap \mbox{\sf P}_{\frac{1}{2}}^+)$. It follows $D\mbox{\sf P}_0^-=G_{-\frac{1}{2}}G_{-\frac{1}{2}} \mbox{\sf P}_0^-\subset G_{-\frac{1}{2}}(G_{-\frac{1}{2}}\mbox{\sf P}_0 \cap \mbox{\sf P}_{\frac{1}{2}}^+)\subset ( G_{-\frac{1}{2}}G_{-\frac{1}{2}}\mbox{\sf P}_0\cap G_{-\frac{1}{2}}\mbox{\sf P}_{\frac{1}{2}}^+) =D\mbox{\sf P}_0\cap G_{-\frac{1}{2}}\mbox{\sf P}_{\frac{1}{2}}^+$. On the other hand $G_{-\frac{1}{2}}\mbox{\sf P}_{\frac{1}{2}}^+\cap D\mbox{\sf P}_0=
 G_{-\frac{1}{2}}\mbox{\sf P}_{\frac{1}{2}}^+\cap 
 (D\mbox{\sf P}_0^+\oplus D\mbox{\sf P}_0^-)=
 G_{-\frac{1}{2}}\mbox{\sf P}_{\frac{1}{2}}^+\cap 
D\mbox{\sf P}_0^- \subset D\mbox{\sf P}_0^-$ since
$D\mbox{\sf P}_0^{\pm}\subset V^{\pm}$ and $G_{-\frac{1}{2}}\mbox{\sf P}_{\frac{1}{2}}^+\subset V^-$. 
\hspace*{\fill} $\Box$\\
$G^+=G_{-\frac{1}{2}}\mbox{\sf P}_{\frac{1}{2}}^+/D\mbox{\sf P}_0^-$ is the
Lie algebra of states invariant under the GSO-projection. 

We will need two further propositions. The first one is the following
\begin{pp1}
There is a symmetric invariant bilinear form $(\, , \, )$ on $G$ satisfying
\begin{enumerate}
\item The adjoint of $k(n),\, k\in h,\, n\in {\mathbb Z}$ is given by $-k(-n)$.
\item The adjoint of $\zeta(r),\, \zeta\in A,\, r\in {\mathbb Z}+\frac{1}{2}$
      is $-\zeta(-r)$.
\item $(e^k,e^l)=-\delta_{k+l,0}$ for $k,l\in L$.
\end{enumerate}
\end{pp1}
{\it Proof:} We will construct an invariant form on $V$ that projects down to an invariant form on $G$ with the above properties. \\
Since $L_1$ is locally nilpotent on $V_{ferm}$ and on $V_{bos}$ it is locally nilpotent on $V=V_{ferm} \otimes V_{bos}$ and we can apply the results of section (\ref{inff}). $V_1^0$ is generated by elements of the form $k(-1)+\zeta (-\frac{1}{2})\xi (-\frac{1}{2}),\, k\in h,\, \zeta,\xi\in A$, so that $(L_1V_1)^0=L_1V_1^0=0$ and ${\mathbb R}1\cap L_1V_1={\mathbb R}1\cap (L_1V_1)^0=0$. 
Hence the dual of $V_0/L_1V_1$ is nontrivial. Define a linear functional on $V_0/L_1V_1$ by putting $f(1+L_1V_1)=-1$ and zero on some complement of $L_1V_1$. We have seen that $f$ induces an invariant form 
$(\, ,\, )$ on $V$ depending on a constant $\lambda$. We will choose $\lambda =i$ and perform the following calculations in the complexification $V_{\mathbb C}={\mathbb C}\otimes_{\mathbb R}V$. Note that this bilinear form is real valued on $V_{\ov{0}}$. Recall the definition of the opposite vertex operator
\[ v_n^*=i^{-2p} \sum_{m\geq 0} \left( \frac{L_1^m}{m!}v \right)_{2p-n-m-2} \]
for $v\in V_p$. For a quasiprimary state $v\in V_p$ this reduces to 
\[ v_n^*=i^{-2p} v_{2p-n-2}\, . \] For example 
\[ L_n^*=\omega_{n+1}^*=\omega_{-n+1}=L_{-n}\, .\]
Let $k\in h$. Then $k(-1)$ is quasiprimary and $k(n)^*=k(-1)^*_n
=-k(-1)_{2-n-2}\linebreak =-k(-n)$  since $k(n)=k(-1)_n$. Analogous we find $\zeta(r)^*=-i\zeta(-r)$ and $(e^k)^*_{-1}=i^{-\la k,k \ra}(e^k)_{\la k,k \ra-1}$ for $\zeta\in A,\, r\in {\mathbb Z}+\frac{1}{2}$ and $k\in L$. Next we show that $(e^k,e^l)=0$ if $k+l\neq 0$. First let $\la k,k+l \ra =0$. Then \beann
(e^k,e^l) &=& ((e^k)_{-1}1,e^l) \\
          &=& (1,(e^k)^*_{-1}e^l) \\
          &=& i^{-\la k,k \ra} (1,(e^k)_{\la k,k \ra-1}e^l) \\
          &=& \varepsilon(k,l)i^{-\la k,k \ra}
                 (1,S_{-\la k,k+l \ra}(k)e^{k+l})\\
          &=& \varepsilon(k,l)i^{-\la k,k \ra}(1,e^{k+l})\\
          &=& \varepsilon(k,l)i^{-\la k,k \ra}f(\pi(e^{k+l}))\\
          &=& 0   \, .
\eeann
If $\la k,k+l \ra \neq 0$ then $\la k,k \ra (e^k,e^l)=(k(0)e^k,e^l)=-(e^k,k(0)e^l)=-\la k,l \ra (e^k,e^l)$ so that $\la k,k+l \ra (e^k,e^l)=0$ and $(e^k,e^l)=0$. Finally 
\beann
(e^k,e^{-k}) 
&=& \varepsilon(k,-k)i^{-\la k,k \ra}(1,S_0(k)e^{0}) \\
&=& \varepsilon(k,-k)i^{-\la k,k \ra}(1,1) \\
&=& -\varepsilon(k,-k)i^{-\la k,k \ra}
\eeann
since $(1,1)=f(\pi(1_{-1}^{*}1))=f(\pi(1))=f(1+L_1V_1)=-1$.\\
If we compensate the sign convention in the definition \ref{surfen} of an invariant form in the definition of the adjoint operator then it is easy to check that the restriction of $(\, , \, )$ to $V_{\ov{0}}$ has the properties stated in the proposition. We are done when we have shown that $(\, ,\, )$ gives a well-defined form on $G$. Let $u\in P_1$ and $v\in P_0$. Then $(u,Dv)=(L_1u,v)=0$ as desired.
\hspace*{\fill} $\Box$\\

The proposition implies that
\be (G^{\alpha},G^{\beta})=0 \quad\mbox{if}\quad \alpha + \beta \neq 0 \, .\ee 

Define an involution $\theta$ of $V_{\ov{0}}$ by $\theta(e^r)=e^{-r},\, 
\theta(k(-n))=-k(-n)$ and $\theta(\zeta(-r))=-\zeta(-r)$. Since $(\, ,\, )$ is $\theta$-invariant the bilinear form $(\, ,\, )_0$ defined by $(x,y)_0=-
(x,\theta(y))$ for $x,y\in V_{\ov{0}}$ is a symmetric bilinear form on 
$V_{\ov{0}}$. It satisfies $(e^k,e^l)_0=\delta_{k,l}$ for $k,l\in L$. The adjoint of $k(n),\, k\in h,\, n\in {\mathbb Z}$ with respect to this form is given by $k(-n)$ and the adjoint of $\zeta(r),\, 
\zeta\in A,\, r\in {\mathbb Z}+\frac{1}{2}$ is $\zeta(-r)$. Note that $(\, ,\, )$ and $(\, ,\, )_0$ have the same kernel so that $(\, ,\, )_0$ gives us another symmetric bilinear form on $G$. It is the bilinear form we know from string theory. If $G_{-\frac{1}{2}}x$ and $G_{-\frac{1}{2}}y$ with $x,y \in \mbox{\sf P}_{\frac{1}{2}}$ are the representatives of two elements in $G$ then $(G_{-\frac{1}{2}}x,G_{-\frac{1}{2}}y)_0=(x,G_{\frac{1}{2}}G_{-\frac{1}{2}}y)_0=(x,[G_{\frac{1}{2}}G_{-\frac{1}{2}}]_+y)_0=2(x,L_0y)_0=(x,y)_0\,$ .
 
The vector space $G_{-\frac{1}{2}}A(-\frac{1}{2})=\{\zeta(-1) | \zeta \in A \}$ is isomorphic to \mbox{$L_{\mathbb R}=L\otimes_{\mathbb Z}{\mathbb R}$.} 
Let $H=G_{-\frac{1}{2}}A(-\frac{1}{2})\Big/ \Big( G_{-\frac{1}{2}}A(-\frac{1}{2})\cap D\mbox{\sf P}_0\Big)$ be the image of $G_{-\frac{1}{2}}A(-\frac{1}{2})$ in $G$. Since 
$G_{-\frac{1}{2}}A(-\frac{1}{2})\subset V^0$ and $(D\mbox{\sf P}_0)^0=D\mbox{\sf P}_0^0=D\{ {\mathbb R}1\otimes e^0\}=0$ we have $G_{-\frac{1}{2}}A(-\frac{1}{2})\cap D\mbox{\sf P}_0 = G_{-\frac{1}{2}}A(-\frac{1}{2})\cap D\mbox{\sf P}_0^0 = 0$ so that $H$ is still isomorphic to $L_{\mathbb R}$. We have already seen that $H$ is an abelian subalgebra of $G$. The restriction of $(\, ,\, )$ to $H$ is equal to $\la \, ,\, \ra$. From $\mbox{\sf P}_{\frac{1}{2}}^0=A(-\frac{1}{2})$ and $D\mbox{\sf P}_0^0 = 0$ we get $G^0
=(G_{-\frac{1}{2}}\mbox{\sf P}_{\frac{1}{2}})^0/(D\mbox{\sf P}_0)^0
=G_{-\frac{1}{2}}\mbox{\sf P}_{\frac{1}{2}}^0/D\mbox{\sf P}_0^0
=G_{-\frac{1}{2}}A(-\frac{1}{2})
=H$.
We will denote an element in $H$ by the corresponding element in 
$L_{\mathbb R}$. 
\begin{pp1} \label{sp}
Let $v\in G^{\alpha}$ be an element of degree $\alpha\in L$. Then 
\be [h,v]=(h,\alpha )v \quad\mbox{for all }h\in H\, .\ee
\end{pp1}
{\it Proof:} $v\in G^{\alpha}$ can be represented by
$G_{-\frac{1}{2}}\big( a\otimes (b\otimes e^{\alpha})\big)$ where $a\in V_{ferm}$ and $b\otimes e^{\alpha}\in V_{bos}$. First suppose that $a$ is homogeneous. Then
\beann
\big( G_{-\frac{1}{2}}h(-{\textstyle \frac{1}{2}}) \big)_0 \big(a\otimes (b\otimes e^{\alpha})\big) 
&=&
\big( 1\otimes (h(-1)\otimes 1)\big)_0\big(a\otimes (b\otimes e^{\alpha})
                                                                     \big)\\
&=& 
\sum_{n\in \mathbb Z} 1_na\otimes (h(-1)\otimes e^0)_{-n-1}(b\otimes 
e^{\alpha}) \\
&=&
a\otimes (h(-1)\otimes e^0)_0(b\otimes e^{\alpha}) \\
&=& a\otimes \la h, \alpha \ra (b\otimes e^{\alpha})\, . 
\eeann
By linearity this holds for arbitrary $a$. Thus
\beann
{[}h,v{]} &=&
\big( G_{-\frac{1}{2}}h(-{\textstyle \frac{1}{2}}) \big)_0
\big( G_{-\frac{1}{2}}\big( a\otimes (b\otimes e^{\alpha})\big)\big)\\
&=&
G_{-\frac{1}{2}} \Big( \big( G_{-\frac{1}{2}}h(-{\textstyle \frac{1}{2}}) \big)_0 \big(a\otimes (b\otimes e^{\alpha})\big)\Big) \\
&=& (h,\alpha)v \, .
\eeann

\vspace{-0.9cm}\hspace*{\fill} $\Box$\\

Note that the last two propositions also hold for the subalgebra $G^+$ of $G$.

\section{The Lorentzian case}

In this section we will consider the special case that $L$ is a Lorentzian lattice. It is of special interest since then the quotient of $G$ resp. $G^+$ by the kernel of the invariant form constructed above is a generalized Kac-Moody algebra which is not true for arbitrary lattices.

Let us assume from now on that $L$ is a Lorentzian lattice.

We will use the following characterization of generalized Kac-Moody algebras due to R. E. Borcherds (cf. \cite{Bo5}).
\begin{thp1} \label{cgk}
Any Lie algebra G satisfying the following five conditions is a generalized 
Kac-Moody algebra.
\samepage{
\begin{enumerate}
\item G has a nonsingular invariant symmetric bilinear form. 
\item G has a selfcentralizing subalgebra H such that G is the sum of the eigenspaces of H and all the eigenspaces are finite-dimensional.
\item The bilinear form restricted to H is Lorentzian.
\item The norms of the roots of G are bounded above.
\item If two roots are positive multiples of the same norm 0 vector then their root spaces commute.
\end{enumerate} }
\end{thp1}

Let ${\cal I}$ be the kernel of the bilinear form $(\, ,\, )$ constructed above. Since $(\, ,\, )$ is invariant ${\cal I}$ is an ideal in $G$. 
We will show that the quotient 
\be {\ov G}=G/{\cal I} \ee
satisfies the five conditions of the theorem.
Note that ${\ov G}$ is isomorphic to $G_{-\frac{1}{2}}\mbox{\sf P}_{\frac{1}{2}}/{\cal I}$ if we denote the kernel of $(\, ,\, )$ on $G_{-\frac{1}{2}}\mbox{\sf P}_{\frac{1}{2}}$ also by ${\cal I}$.
${\cal I}$ is $L$-graded, i.e.
\be {\cal I}=\bigoplus_{\alpha \in L}{\cal I}^{\alpha}\quad\mbox{with}\quad 
    {\cal I}^{\alpha}={\cal I}\cap G^{\alpha}\, . \ee
In order to prove this let $x=\sum_{i\in I}x^{\alpha_i}$ with $x^{\alpha_i}\in 
G^{\alpha_i}$ be in ${\cal I}$. We have to show that $x^{\alpha_j},\, j\in I$, is already in ${\cal I}$. Since $(G^{\alpha},x^{\alpha_j})=0$ for all 
$\alpha\neq -\alpha_j $ it is enough to show that 
$(G^{-\alpha_j},x^{\alpha_j})=0$. But this follows from $0=(G^{-\alpha_j},x)=
\sum_{i\in I}(G^{-\alpha_j},x^{\alpha_i})=(G^{-\alpha_j},x^{\alpha_j})$.

Thus ${\ov G}$ is also naturally $L$-graded
\be {\ov G}=\bigoplus_{\alpha \in L}{\ov G}^{\alpha} 
     \quad\mbox{with}\quad
    {\ov G}^{\alpha}=G^{\alpha}/{\cal I}^{\alpha} \ee
and 
\be \left[{\ov G}^{\alpha},{\ov G}^{\beta}\right] \subset {\ov G}^{\alpha + \beta}\, . \ee
The vector spaces ${\ov G}^{\alpha}$ are finite-dimensional by construction. 
${\ov G}^{\alpha}$ and ${\ov G}^{-\alpha}$ are isomorphic.

An element $\alpha\in L\backslash \{0\}$ for which ${\ov G}^{\alpha}$ is nontrivial is called 
root of ${\ov G}$ and the corresponding ${\ov G}^{\alpha}$ is called root 
space. The norm of $\alpha$ is given by $(\alpha,\alpha)=\alpha^2$. 

$(\, , \, )$ pairs nondegenerately ${\ov G}^{\alpha}$ and ${\ov G}^{-\alpha}$.
Since ${\ov G}^{\alpha}$ and ${\ov G}^{-\alpha}$ are of the same finite dimension this pairing is even nonsingular. Thus ${\ov G}$ satisfies the first condition of Theorem (\ref{cgk}).

The image ${\ov H}$ of $H$ in ${\ov G}$ is isomorphic to $H$ since 
$(\, , \, )$ is nondegenerate on $H$. Furthermore ${\ov H}={\ov G}^0$ and 
\be {\ov G}={\ov H}\oplus \bigoplus_{\alpha \neq 0}{\ov G}^{\alpha}. \ee
The spaces ${\ov G}^{\alpha}$ are finite-dimensional eigenspaces of 
${\ov H}$ by Proposition (\ref{sp}). We now show that ${\ov H}$ 
is selfcentralizing. Let $x=h+\sum_{i\in I}x^{\alpha_i}$ with $x^{\alpha_i}\in 
{\ov G}^{\alpha_i},\, {\alpha_i}\neq 0$ and $h\in {\ov H}$ be in the centralizer of ${\ov H}$. Then $0=[h',x]=\sum_{i\in I}
[h',x^{\alpha_i}]=\sum_{i\in I}(h',{\alpha_i})x^{\alpha_i}$ for all 
$h'\in {\ov H}$. If some $x^{\alpha_j}$ are nonzero they are linearly 
independent and $(h',{\alpha_j})=0$ for all $h'\in {\ov H}$. 
But this is impossible by the nondegeneracy of $(\, ,\, )$.
Thus all $x^{\alpha_i}$ are zero and $x$ is in ${\ov H}$.

Clearly the restriction of $(\, ,\, )$ to ${\ov H}$ is Lorentzian since $\la 
\, , \, \ra$ is.

The norms of the roots are integer and bounded above by 1.

Finally we have
\begin{lp1}
Let two roots be positive multiples of the same norm 0 vector then their root spaces commute.
\end{lp1}
{\it Proof:} Let $r\in L$ be of norm zero, i.e. $(r,r)=0$, and let $x\in {\ov G}^{mr}$ and $y\in {\ov G}^{nr}$ with $m,n > 0$ be nonzero. Then 
$x=G_{-\frac{1}{2}}a(-\frac{1}{2})e^{mr}$ and 
$y=G_{-\frac{1}{2}}b(-\frac{1}{2})e^{nr}$ modulo ${\cal I}$. The physical state condition implies $(a,r)=(b,r)=0$.
\beann
\lefteqn{\big( G_{{\textstyle -\frac{1}{2}}}a({\textstyle -\frac{1}{2}})e^{mr} \big)_0
         \big( b({\textstyle -\frac{1}{2}})e^{nr} \big) } \\ 
&=& 
\big( (a(-1)-ma({\textstyle -\frac{1}{2}})r({\textstyle -\frac{1}{2}}))e^{mr} \big)_0 
\big( b({\textstyle -\frac{1}{2}})e^{nr} \big) \\
&=&
(a(-1)e^{mr})_0(b({\textstyle -\frac{1}{2}})e^{nr})
-m(a({\textstyle -\frac{1}{2}})r({\textstyle -\frac{1}{2}})e^{mr})_0(b({\textstyle -\frac{1}{2}})e^{nr}) \\
&=&
\sum_{i\in \mathbb Z} 1_ib({\textstyle -\frac{1}{2}})\otimes (a(-1)e^{mr})_{-i-1}(e^{nr})
\\
& &
-m \sum_{i\in \mathbb Z} a({\textstyle -\frac{1}{2}})r({\textstyle -\frac{1}{2}})_ib({\textstyle -\frac{1}{2}})
\otimes (e^{mr})_{-i-1}(e^{nr}) \\
&=& 
b({\textstyle -\frac{1}{2}})\otimes (a(-1)e^{mr})_0(e^{nr}) \\
& &
-m \sum_{i\geq 0}a({\textstyle -\frac{1}{2}})r({\textstyle -\frac{1}{2}})_ib({\textstyle -\frac{1}{2}})\otimes 
\varepsilon(mr,nr)S_i(mr)e^{(m+n)r}
\eeann
Using (\ref{sgg}) we find
\[ (a(-1)e^{mr})_0(e^{nr})=0 \, .\]
The products $a({\textstyle -\frac{1}{2}})r({\textstyle -\frac{1}{2}})_ib({\textstyle -\frac{1}{2}})$ can be calculated by (\ref{schellistdoof}). Terms with 
$i \geq 1$ vanish and 
\[ a({\textstyle -\frac{1}{2}})r({\textstyle -\frac{1}{2}})_0b({\textstyle -\frac{1}{2}})=-(a,b)r({\textstyle -\frac{1}{2}}). \]
Thus 
\beann
\big( G_{-\frac{1}{2}}a({\textstyle  -\frac{1}{2}})e^{mr} \big)_0
         \big( b({\textstyle -\frac{1}{2}})e^{nr} \big)   
&=&
m\varepsilon(mr,nr)(a,b)r({\textstyle -\frac{1}{2}})e^{(m+n)r} \\
&=&
{\textstyle \frac{m}{m+n}} \varepsilon(mr,nr) (a,b)G_{-\frac{1}{2}}e^{(m+n)r}.
\eeann
It is easy to check that 
\[ L_0e^{(m+n)r}=G_{\frac{1}{2}}e^{(m+n)r}=G_{\frac{3}{2}}e^{(m+n)r}=0 \]
so that $e^{(m+n)r} \in \mbox{\sf P}_0$. Hence
\beann
{[}x,y{]} 
&=&
\big( G_{{\textstyle -\frac{1}{2}}}a({\textstyle -\frac{1}{2}})e^{mr} \big)_0
\big( G_{{\textstyle -\frac{1}{2}}}b({\textstyle -\frac{1}{2}})e^{nr} \big) \\
&=&
G_{-\frac{1}{2}}\big( \big(
G_{-\frac{1}{2}}a({\textstyle  -\frac{1}{2}})e^{mr} \big)_0
          \big( b({\textstyle  -\frac{1}{2}})e^{nr} \big) \big) \\
&=& {\textstyle \frac{m}{m+n}} \varepsilon(mr,nr) (a,b)De^{(m+n)r} \\
&=& 0 \quad (\, \mbox{mod}\; D\mbox{\sf P}_0\, )
\eeann
\hspace*{\fill} $\Box$\\

We have proven the following
\begin{thp1}
${\ov G}$ is a generalized Kac-Moody algebra. The Cartan subalgebra ${\ov H}$
is isomorphic to $L\otimes_{\mathbb Z}{\mathbb R}$. The roots have integer 
norms bounded above by $1$.
\end{thp1}

The same arguments hold for ${\ov G}^+=G^+/{\cal I}$. We can be more precise on the norms of the roots. The roots of ${\ov G}^+$ have even norms bounded above by $0$. Thus
\begin{thp1}
${\ov G}^+$ is a generalized Kac-Moody algebra with Cartan subalgebra ${\ov H}$
isomorphic to $L\otimes_{\mathbb Z}{\mathbb R}$. ${\ov G}^+$ has only imaginary roots of even norm.
\end{thp1}

In the case that $L$ is a Lorentzian lattice with rank less or equal to $10$ we can even calculate the multiplicities of the roots. 

First let the rank $d$ of $L$ be smaller than or equal to $9$. Then by the 
"no-ghost-theorem" 
$(\, ,\, )_0$ is positive definite on $G$ and the kernel of 
$(\, ,\, )$ in $G$ is trivial. For a root $\alpha$ we have 
$\mbox{dim }{\ov G}^{\alpha}= \mbox{dim }{G}^{\alpha}=
\mbox{dim }\mbox{\sf P}_{\frac{1}{2}}^{\alpha}-\mbox{dim }\mbox{\sf P}_0^{\alpha}$ since 
$G_{-\frac{1}{2}}$ is injective on $\mbox{\sf P}_{\frac{1}{2}}$ and $D$ is injective on $\mbox{\sf P}_0^{\alpha}$. 
Thus
\be \mbox{dim }{\ov G}^{\alpha}= c_{d-1}({\textstyle \frac{1}{2}}(1-\alpha^2))-
                c_{d-1}(-{\textstyle \frac{1}{2}} \alpha^2) \ee
by formula (\ref{ahd}).
The same result holds for the roots of ${\ov G}^+$. We summarize this in
\begin{thp1}
Let $L$ be a Lorentzian lattice of rank $2\leq d\leq 9$. Then ${\ov G}$ and ${\ov G}^+$
 are generalized Kac-Moody algebras. The roots of ${\ov G}$ resp. ${\ov G}^+$ are the $\alpha\in L\backslash \{0\}$ with $\alpha^2\leq 1$ resp. $\alpha^2\in 
2{\mathbb Z}$ and $\alpha^2\leq 0$. In both cases the multiplicity of 
a root ${\alpha}$ is given by $c_{d-1}({\textstyle \frac{1}{2}}(1-\alpha^2))-
                c_{d-1}(-{\textstyle \frac{1}{2}} \alpha^2).$
\end{thp1}

If $L$ is a 10-dimensional Lorentzian lattice then the kernel of 
$(\, ,\, )$ is nontrivial. Denote the kernel of $(\, ,\, )_0$ by ${\cal I}_0$.
Then ${\cal I}={\cal I}_0$ and $\mbox{dim }{\ov G}^{\alpha}=\mbox{dim }
\mbox{\sf P}_{\frac{1}{2}}^{\alpha}/{\cal I}_0^{\alpha}$. 
\begin{thp1}
Let $L$ be a 10-dimensional Lorentzian lattice. Then ${\ov G}$ and ${\ov G}^+$
 are generalized Kac-Moody algebras. The roots of ${\ov G}$ resp. ${\ov G}^+$ are the $\alpha\in L\backslash \{0\}$ with $\alpha^2\leq 1$ resp. $\alpha^2\in 
2{\mathbb Z}$ and $\alpha^2\leq 0$. In both cases the multiplicity of 
${\alpha}$ is given by $c_8(\frac{1}{2}(1-{\alpha}^2))$.
\end{thp1}

The following table shows that the multiplicities are indeed equal to the number of states of the corresponding mass level.
\[ \renewcommand{\arraystretch}{1.2}
\begin{array}{r|c}
{\rule[-1,5ex]{0ex}{1,2ex}}
\alpha^2 & c_8(\frac{1}{2}(1-{\alpha}^2)) \\ \hline
\, 1   \,     & 1     \\
\, 0   \,     & 8     \\
\, -1  \,     & 36    \\
\, -2  \,     & 128   \\
\, -3  \,     & 402
\end{array} \]

The asymptotic behaviour of $c_8(n)$ can by calculated by the methods described in \cite{Bo6} or \cite{GSW} vol.1 p.118. We find
\be  c_8(n)\sim {\textstyle \frac{1}{2}}n^{-\frac{11}{4}}e^{2\pi\sqrt{n}}\, .\ee
Note that if $\alpha^2$ is even then $c_8(\frac{1}{2}(1-{\alpha}^2))$ is equal to the coefficient of $q^{-\frac{1}{2}\alpha^2}$ in 
\[ 8 \prod_{m=1}^{\infty}\left( 
                 \frac{1+q^m}{1-q^m}
                                      \right)^8 \]
by Jacobi's "rather obscure formula".

\chapter{Outlook}

We have seen above that the Fock space of the compactified Neveu-Schwarz superstring carries the structure of a vertex superalgebra and that the physical states form a Lie algebra. We will denote them in this chapter by $V_{\ov 0}$ resp. $G_{\ov 0}$. 

From the physical point of view one should expect that $G_{\ov 0}$ is the even part of a superalgebra representing the full superstring moving on a torus. 

Although this idea seems quite simple its mathematical realization is rather 
difficult. 

We make the following conjectures:\\
1) The Ramond sector $V_{\ov 1}$ is a twisted module for the vertex superalgebra $V_{\ov 0}$ of the compactified Neveu-Schwarz superstring. This structure induces a $G_{\ov 0}$-module structure of some quotient $G_{\ov 1}$ in the Ramond sector.\\
2) There is an intertwiner mapping two states from the Ramond sector into the Neveu-Schwarz sector. In physics this map is called fermion-emission vertex. The problem is to find a mathematically correct description of this operator in the formalism of vertex superalgebras. The intertwiner induces then a map $G_{\ov 1}\times G_{\ov 1}\rightarrow G_{\ov 0}$.\\
The conjectures imply that $G=G_{\ov 0}\oplus G_{\ov 1}$ is a Lie superalgebra. It is the Lie superalgebra of a superstring moving on a torus.

Once that $G$ is constructed one can make similar considerations as for the Neveu-Schwarz superstring. They lead to the first realization of generalized Kac-Moody superalgebras! One can determine the roots and their multiplicities by modifying the arguments given in this work. 

Furthermore there is some evidence from the theory of modular forms that it is even possible to compute the simple roots of the Lie superalgebra of physical states if the superstring moves on the torus ${\mathbb R}^{9,1}/ I{\!}I_{9,1}$. This would give a fake monster Lie superalgebra.
Unfortunately, similar arguments do not seem to apply to the case of the Neveu-Schwarz superstring.

\begin{appendix}

\chapter[Physical subspaces]{The dimensions of the physical subspaces}

In this chapter we calculate the dimensions of the spaces $P^{\alpha}_n$ resp. 
$\mbox{\sf P}^{\alpha}_n$. Some arguments are similar to those of Goddard and Thorne in their proof of the "no-ghost-theorem" (cf. \cite{GT}). 

\section{The bosonic case}

Let $V$ be the vertex algebra of an integral nondegenerate lattice $L$ as defined in section \ref{bosc}. $V$ is the Fock space of a compactified bosonic string. For ${\alpha}\in L$ let 
$V^{\alpha}=S(\hat{h}^-)\otimes e^{\alpha}$ be the ${\alpha}$-subspace of $V$. It is a module for the Virasoro algebra. We define a bilinear form $\la \, , \, \ra$ on $V^{\alpha}$ by putting $h(-m)^{\dagger}=h(m)$ and $\la e^{\alpha},e^{\alpha}\ra=1$. This form is nondegenerate since the bilinear form on $L$ is nondegenerate. The adjoint of the Virasoro operator $L_m$ with respect to $\la \, , \, \ra$ is given by $L_m^{\dagger}=L_{-m}$. The spaces $V^{\alpha}_n=\{ v\in V^{\alpha}\, | \, L_0v=nv \}$ are orthogonal for different $n$ and $V^{\alpha}=\oplus_{n \in {\frac{1}{2}} {\mathbb Z}}V^{\alpha}_n$. It follows that the spaces $V^{\alpha}_n$ are nondegenerate. Define $P^{\alpha}_n=\{v\in V^{\alpha}_n\, | \, L_mv=0 \, \mbox{for } m>0 \}$. For the rest of the section we fix an ${\alpha}\neq 0$. Choose a vector $k$ in $L_{\mathbb R}$ that is linear independent from ${\alpha}$ and satisfies $\la {\alpha},k \ra \neq 0$. If $L_{\mathbb R}$ has nonzero Witt index we impose the additional condition $k^2=0$. The existence of such a vector $k$ can be proved by using Witt's theorem. Define operators $K_m=k(m),\, m\in {\mathbb Z}$. Then 
\bea 
{[}K_m,K_n{]}  &=& m\delta_{m+n,0}k^2 \label{aagl1}\\
{[}K_m,L_n{]} &=& mK_{m+n} \label{aagl2} \, .
\eea
Let $T^{\alpha}$ be the subspace of $P^{\alpha}$ annihilated by all $K_m,\, m>0$. It can be interpreted as the space of tranversal states with momentum ${\alpha}$. The commutation relations imply $T^{\alpha}=\{ v\in V^{\alpha}\, | \, L_1v=L_2v=K_1v=0 \}$. $L_0$ defines a grading on $T^{\alpha}$, i.e. $T^{\alpha}=\oplus_{n \in {\frac{1}{2}} {\mathbb Z}}T^{\alpha}_n$ with 
$T^{\alpha}_n=\{v\in P^{\alpha}_n\, | \, K_mv=0 \; \mbox{for } m>0 \}$. 
\begin{l1} \label{aala}
Let $v\in T^{\alpha}_{M+\frac{1}{2}{\alpha}^2},\, M\in {\mathbb Z}$ with $v\neq 0$. Then the vectors 
\be L^{\lambda_1}_{-1}\ldots L^{\lambda_n}_{-n}
    K^{\mu_1}_{-1}\ldots K^{\mu_m}_{-m} v \label{agaga} \ee
are linearly independent.
\end{l1}
{\it Proof}: cf. \cite{S} p. 137.

If $v$ and $w$ are orthogonal vectors in $T^{\alpha}_{M+\frac{1}{2}{\alpha}^2}$ then the vectors obtained by acting with $L$'s and $K$'s on them are clearly still orthogonal.
 
\begin{ll1} \label{aasurf}
Let $v\in T^{\alpha}_{M+\frac{1}{2}{\alpha}^2},\, M\in {\mathbb Z}$ with $v^2\neq 0$. Then the vectors 
\be L^{\lambda_1}_{-1}\ldots L^{\lambda_n}_{-n}
    K^{\mu_1}_{-1}\ldots K^{\mu_m}_{-m} v \label{tuberider} \ee
with $\sum r\lambda_r + \sum s\mu_s >0$ span a nondegenerate subspace of $V^{\alpha}$. 
\end{ll1}
{\it Proof}: This is trivially true if $L_{\mathbb R}$ has Witt index zero. In the other case the vectors (\ref{tuberider}) can be ordered so that the matrix of inner products is upper triangular with nonzero diagonal elements (cf. \cite{GSW} p. 104). The conditions $\la {\alpha},k \ra\neq 0$ and $k^2=0$ enter into the proof. \hspace*{\fill} $\Box$\\

If $v$ is an element of the subspace generated by the elements of the form     (\ref{tuberider}) satisfying $L_nv=K_nv=0$ for all $n>0$ then 
$v=0$ by the nondegeneracy of $\la \, ,\, \ra$. 

We recall a simple fact from linear algebra.
\begin{ll1}
Let $E$ be a finite dimensional real vectorspace with a nondegenerate symmetric bilinear form and $F$ be subspace of $E$. Then the following conditions are equivalent: 
\begin{enumerate}
\item $F$ is nondegenerate
\item the space $F^{\bot}$ of vectors orthogonal to $F$ is nondegenerate
\item $E=F\oplus F^{\bot}$ 
\end{enumerate}
\end{ll1}  
Now we prove
\begin{ll1}
Let $N\geq 0$. Then $T^{\alpha}_{N+\frac{1}{2}{\alpha}^2}$ is nondegenerate and the vector 
space generated by 
\[  L^{\lambda_1}_{-1}\ldots L^{\lambda_n}_{-n}
    K^{\mu_1}_{-1}\ldots K^{\mu_m}_{-m} v \]
with $v\in T^{\alpha}_{M+\frac{1}{2}{\alpha}^2}$ such that $v^2\neq 0$ and $\sum r\lambda_r + \sum s\mu_s + M=N$ is $V^{\alpha}_{N+\frac{1}{2}{\alpha}^2}$.
\end{ll1}
{\it Proof}: 
Define $G^{\alpha}_{N+\frac{1}{2}{\alpha}^2}$ for $N>0$ as the vector space generated by the elements $L^{\lambda_1}_{-1}\ldots L^{\lambda_n}_{-n}
    K^{\mu_1}_{-1}\ldots K^{\mu_m}_{-m} v$ with $v\in T^{\alpha}_{M+\frac{1}{2}{\alpha}^2}, \, v^2\neq 0$ and $\sum r\lambda_r + \sum s\mu_s + M=N$ where $M$ runs from $0$ to $N-1$. Put $G^{\alpha}_{N+\frac{1}{2}{\alpha}^2}=0$ for $N=0$. We prove by induction that 
\be 
V^{\alpha}_{N+\frac{1}{2}{\alpha}^2}=T^{\alpha}_{N+\frac{1}{2}{\alpha}^2}\oplus G^{\alpha}_{N+\frac{1}{2}{\alpha}^2}
\ee
and
\be
\left( T^{\alpha}_{N+\frac{1}{2}{\alpha}^2}\right)^{\bot}=G^{\alpha}_{N+\frac{1}{2}{\alpha}^2}\, .
\ee
The statement then follows from the previous lemma.
It is easy to check the case $N=0$.
Let $w\in V^{\alpha}_{1+\frac{1}{2}{\alpha}^2}$. Then $w=s(-1)\otimes e^{\alpha}$. The conditions 
$L_nv=K_nv=0$ for $n>0$ are equivalent to $\la s,r \ra =\la s,k \ra =0$. Hence
$T^{\alpha}_{1+\frac{1}{2}{\alpha}^2}=\{ s(-1)\otimes e^{\alpha} \, | \, s\bot r\, \mbox{and}\, 
s\bot k \}$ and $V^{\alpha}_{1+\frac{1}{2}{\alpha}^2}=T^{\alpha}_{1+\frac{1}{2}{\alpha}^2}\oplus ( T^{\alpha}_{1+\frac{1}{2}{\alpha}^2})^{\bot}$. The elements $L_{-1}e^{\alpha}=r(-1)\otimes e^{\alpha}$ and $K_{-1}e^{\alpha}=k(-1) \otimes e^{\alpha}$ span $( T^{\alpha}_{1+\frac{1}{2}{\alpha}^2})^{\bot}=G^{\alpha}_{1+\frac{1}{2}{\alpha}^2}$. 
Since $T^{\alpha}_{1+\frac{1}{2}{\alpha}^2}$ is nondegenerate it possesses a basis consisting of elements with nonzero norm. This proves the assertions for $N=1$. 
Now assume that they are true for $N'<N$. Then 
\beann
L_{-1}V^{\alpha}_{(N-1)+\frac{1}{2}{\alpha}^2} &\subset & G^{\alpha}_{N+\frac{1}{2}{\alpha}^2}\\
L_{-2}V^{\alpha}_{(N-2)+\frac{1}{2}{\alpha}^2} &\subset & G^{\alpha}_{N+\frac{1}{2}{\alpha}^2}
\eeann
and 
\beann
K_{-1}V^{\alpha}_{(N-1)+\frac{1}{2}{\alpha}^2} &\subset & G^{\alpha}_{N+\frac{1}{2}{\alpha}^2}\, .
\eeann
Let $v\in V^{\alpha}_{N+\frac{1}{2}{\alpha}^2}$ be orthogonal to $G^{\alpha}_{N+\frac{1}{2}{\alpha}^2}$.
Then $\la v,L_{-1}V^{\alpha}_{(N-1)+\frac{1}{2}{\alpha}^2}\ra=\la L_1v,V^{\alpha}_{(N-1)+\frac{1}{2}{\alpha}^2}\ra=0$ and $L_1v=0$ since $V^{\alpha}_{(N-1)+\frac{1}{2}{\alpha}^2}$ is nondegenerate. In the same way $L_2v=K_1v=0$. Hence $v\in T^{\alpha}_{N+\frac{1}{2}{\alpha}^2}$ and $(G^{\alpha}_{N+\frac{1}{2}{\alpha}^2})^{\bot}\subset 
T^{\alpha}_{N+\frac{1}{2}{\alpha}^2}$. Clearly $T^{\alpha}_{N+\frac{1}{2}{\alpha}^2}\subset (G^{\alpha}_{N+\frac{1}{2}{\alpha}^2})^{\bot}$ so that we have equality. $V^{\alpha}_{N+\frac{1}{2}{\alpha}^2}$ is the sum of $G^{\alpha}_{N+\frac{1}{2}{\alpha}^2}$ and $T^{\alpha}_{N+\frac{1}{2}{\alpha}^2}$. By Lemma \ref{aasurf} this sum is direct. Finally it is easy to see that $(T^{\alpha}_{N+\frac{1}{2}{\alpha}^2})^{\bot}= G^{\alpha}_{N+\frac{1}{2}{\alpha}^2}$. \hspace*{\fill} $\Box$\\

The number of partitions $p_d$ into $d$ colours is defined by
\be  
\sum_{n=0}^{\infty} p_d(n)q^n = 
     \prod_{m=1}^{\infty}
     \frac{1}{ \left( 1-q^m \right)^d }  \, . 
\ee
The first coefficients are
\beann 
p_d(0) &=& 1 \\
p_d(1) &=& d \\
p_d(2) &=& 2d+{\textstyle {d \choose 2}} \\
p_d(3) &=& d+2d^2+{\textstyle {d \choose 3}}\, .
\eeann
Let $1\leq c\leq d-1$. Then directly from the definition of $p_d(n)$ we get 
\be p_d(n)= \sum_{k=0}^{n} p_c(n-k)p_{d-c}(n)\, . \ee

We now can calculate the dimensions of the vector spaces 
$V^{\alpha}_{N+\frac{1}{2}{\alpha}^2},T^{\alpha}_{N+\frac{1}{2}{\alpha}^2}$ and $P^{\alpha}_{N+\frac{1}{2}{\alpha}^2}$.
Let $d$ be the rank of the lattice $L$. Then the  dimension of $V^{\alpha}_{N+\frac{1}{2}{\alpha}^2}$ is simply
\be \mbox{dim} \, V^{\alpha}_{N+\frac{1}{2}{\alpha}^2}=p_d(N) \, . \ee
We prove by induction that
\be \mbox{dim} \, T^{\alpha}_{N+\frac{1}{2}{\alpha}^2}=p_{d-2}(N) \, . \label{aabla} \ee
This is obviously true for $N=0$ and $N=1$. $V^{\alpha}_{N+\frac{1}{2}{\alpha}^2}$ is generated by vectors $L^{\lambda_1}_{-1}\ldots L^{\lambda_n}_{-n}
                K^{\mu_1}_{-1}\ldots K^{\mu_m}_{-m} v $,
where $v$ runs through an orthogonal basis of $T^{\alpha}_{M+\frac{1}{2}{\alpha}^2}$ and $\sum r\lambda_r + \sum s\mu_s + M=N$. Hence 
\[ \mbox{dim} \, V^{\alpha}_{N+\frac{1}{2}{\alpha}^2}= \sum_{M=0}^{N}
   \mbox{dim} \, T^{\alpha}_{M+\frac{1}{2}{\alpha}^2}p_2(N-M)\,. \]
Suppose that (\ref{aabla}) is true for $M<N$. Then 
\[ p_d(N)=\mbox{dim} \, V^{\alpha}_{N+\frac{1}{2}{\alpha}^2}=
    \sum_{M=0}^{N}p_{d-2}(M)p_2(N-M) -p_{d-2}(N)+
      \mbox{dim} \, T^{\alpha}_{N+\frac{1}{2}{\alpha}^2}\, . \]
This proves (\ref{aabla}).

Since the spaces $T^{\alpha}_{N+\frac{1}{2}{\alpha}^2}$ are nondegenerate we can choose an orthogonal basis of $T^{\alpha}$ consisting of vectors which are eigenvectors of $L_0$ and 
have nonzero norm. By acting with $L$'s and $K$'s on 
this basis we get an orthogonal basis of $V^{\alpha}$ so that $V^{\alpha}$ is the direct sum of pairwise orthogonal Verma modules for the Virasoro algebra. We write these modules as 
${\mathbb R}v_i\oplus C_i$ where $v_i$ is the highest weight vector. Then the spaces $C_i$ are nondegenerate since $V^{\alpha}$ is. Now let
$v$ be in $P_n^{\alpha}$. We can write $v=\sum (\lambda_iv_i+c_i)$. The condition $L_nv=0$ for $n>0$ implies $L_nc_i=0$ and $\la c_i,C_i\ra =0$. Hence all the $c_i$ are zero. It follows that $v$ is in $P_n^{\alpha}$ if and only if $v$ is the linear combination of some $v_i$ with $L_0v_i=nv_i$. Thus the dimension of $P_n^{\alpha}$ is equal to the number of highest weight vectors $v_i$ with $L_0v_i=nv_i$. Define $N=n-\frac{1}{2}{\alpha}^2$. The highest weight vectors $v_i$ are of the form $K^{\mu_1}_{-1}\ldots K^{\mu_m}_{-m} w$ where $w$ is an element of an orthogonal basis of $T^{\alpha}_{M+\frac{1}{2}{\alpha}^2}$. The condition $L_0v_i=nv_i$ is equivalent to $\sum s\mu_s +M=N$ so that the number of highest weight vectors $v_i$ with $L_0v_i=nv_i$ is given by  
\[ \sum_{M=0}^N \mbox{dim} \, T^{\alpha}_{M+\frac{1}{2}{\alpha}^2}p_1(N-M)=
   \sum_{M=0}^N p_{d-2}(M)p_1(N-M) =p_{d-1}(N)\, . \]
We have proven the main result of this section, namely
\be \mbox{dim} \, P^{\alpha}_n=p_{d-1}(n-{\textstyle \frac{1}{2}}{\alpha}^2)\, . \ee

\section{The Neveu-Schwarz case}  
 
The Neveu-Schwarz case is completely analogous to the bosonic case so that we will only describe the modifications that must be made. 

Let $L$ be an integral nondegenerate lattice and $V$ be the vertex algebra of the compactified Neveu-Schwarz superstring (cf. section \ref{cpnss}). The ${\alpha}$-subspace $V^{\alpha}=\bigoplus_{n\geq 0}\wedge^n (\hat{A}^-)
\otimes S(\hat{h}^-)\otimes e^{\alpha}$ is a module for the Neveu-Schwarz algebra. We define a bilinear form $\la \, ,\, \ra$ on $V^{\alpha} $as above with the additional condition $b(-r)^{\dagger}=b(r)$. On $V_{\ov{0}}$ this form coincides with the bilinear form introduced in section  
\ref{bigwaves}. The adjoints of the Neveu-Schwarz operators are given by $L_m^{\dagger}=L_{-m}$ and $G_r^{\dagger}=G_{-r}$. The physical subspace of conformal weight $n$ and momentum ${\alpha}$ is $\mbox{\sf P}_{n}^{\alpha}=\{v\in V_n^{\alpha}|\, L_mv=0,\, G_rv=0\, \mbox{ for all } m,r> 0 \}$. Let ${\alpha}$ and $k$ be as in the previous section and define operators $H_r=k(r), r\in {\mathbb Z}+\frac{1}{2}$. Apart from (\ref{aagl1}) and (\ref{aagl2}) we get
\[ \renewcommand{\arraystretch}{1.3}
\begin{array}{lcl} 
{[}H_r,H_s{]}   &=& \delta_{r+s,0}k^2 \\
{[}H_r,K_m{]}   &=& 0 \\
{[}H_r,L_n{]}   &=& (r+{\textstyle \frac{n}{2}})H_{r+n}\\
{[}H_r,G_s{]}_+ &=& K_{r+s} \\
{[}K_m,G_r{]}   &=& mH_{m+r}\, . 
\end{array} \]
The tranversal states are the physical states annihilated by all $K_m$ and $H_r$ with $m,r>0$. The commutation relations imply $\mbox{\sf T}^{\alpha}=\{ v\in V^{\alpha}\, | \, G_{\frac{1}{2}}v=G_{\frac{3}{2}}v=H_{\frac{1}{2}}v=0 \}$. 
The lemmas from the previous section hold with the modification that the vectors (\ref{agaga}) must be replaced by vectors of the form
\be G_{-1/2}^{{\varepsilon}_{1/2}}\ldots 
    G_{-r}^{\varepsilon_{r}}
    L^{\lambda_1}_{-1} \ldots  L^{\lambda_n}_{-n}
    H^{{\delta}_1}_{-1}\ldots  H^{{\delta}_s}_{-s}
    K^{\mu_1}_{-1}\ldots K^{\mu_m}_{-m} v\, .
\ee  
The fermionic occupation numbers $\varepsilon_r$ and $\delta_s$ can only take the values zero or one. 

To calculate the dimensions of the vector spaces $V^{\alpha}_{N+\frac{1}{2}{\alpha}^2},\mbox{\sf T}^{\alpha}_{N+\frac{1}{2}{\alpha}^2}$ and $\mbox{\sf P}^{\alpha}_{N+\frac{1}{2}{\alpha}^2}$ we define numbers $c_d(n)$ by
\be  
\sum_{n=0}^{\infty} c_d(n)q^n = 
     \prod_{m=1}^{\infty}
     \left( \frac{1+q^{m-\frac{1}{2}}}{1-q^m} \right)^d   \, . 
\ee
If the rank of the lattice is $d$ then the dimension of $V^{\alpha}_{N+\frac{1}{2}{\alpha}^2}$ is 
\be \mbox{dim} \, V^{\alpha}_{N+\frac{1}{2}{\alpha}^2}=c_d(N) \ee 
and as in the bosonic case one shows that
\be \mbox{dim} \, \mbox{\sf T}^{\alpha}_{N+\frac{1}{2}{\alpha}^2}=c_{d-2}(N) \ee  
and 
\be 
\mbox{dim} \, \mbox{\sf P}^{\alpha}_{N+\frac{1}{2}{\alpha}^2}=c_{d-1}(N)\, .
\ee  

\end{appendix}

\chapter*{Acknowledgments}

I want to thank Prof. H. Nicolai for interesting discussions, inspiring ideas and mental support. 

It is a pleasure to thank Prof. R. E. Borcherds for stimulating discussions
and valuable comments. I am also grateful to Prof. P. Slodowy, Dr. V. Schomerus and Dr. R. W. Gebert for helpful conversations.

There are some other people who have contributed in one or another way to this work. I only want to mention Dr. Y. Xylander and M. Meier-Schellersheim who always cared for a good mood in the study.

Last but not least I thank the University of Hamburg, 
the Richard-Winter-Stiftung
and the Deutsche Forschungsgemeinschaft for financial support.

\end{document}